%% file: main.tex
\documentclass[12pt,letterpaper]{article}
\usepackage[a4paper, total={7in, 10in}]{geometry}

\usepackage{graphicx}
\usepackage{helvet}
\usepackage{authblk}
\usepackage{hyperref}
\usepackage{amsmath} 
\usepackage{amssymb} 
\usepackage{orcidlink} 
\usepackage[super,comma,sort&compress]  
   {natbib}
\usepackage[right]{lineno} \linenumbers

\makeatletter
\renewcommand{\maketitle}{\bgroup\setlength{\parindent}{0pt}
\begin{flushleft}
  \textbf{\@title}
  
  \@author
\end{flushleft}\egroup}
\makeatother

\hypersetup{
  colorlinks   = true, 
  urlcolor     = blue, 
  linkcolor    = blue, 
  citecolor   = blue 
}

\usepackage{verbatim} 

\title
{Tunable supercontinuum in multimode fiber via bending-induced dispersion modification}
\date{}

\author[1,2,\orcidlink{0000-0003-3986-8416}]{Li-Yu Yu}
\author[1,2]{Honghao Cao}
\author[1,2]{Kunzan Liu}
\author[1,2]{Chao Li}
\author[3]{Brandon Weissbourd}
\author[4]{Subhash Kulkarni}
\author[1,2,*,\orcidlink{0000-0002-1243-1815}]{Sixian You}

\affil[1]{Department of Electrical Engineering and Computer Science, Massachusetts Institute of Technology, Cambridge, MA, USA}
\affil[2]{Research Laboratory of Electronics, Massachusetts Institute of Technology, Cambridge, MA, USA}
\affil[3]{Department of Biology, Massachusetts Institute of Technology, Cambridge, MA, USA}
\affil[4]{Beth Israel Deaconess Medical Center, Harvard Medical School, Boston, MA, USA}

\affil[*]{Correspondence: sixian@mit.edu}

\begin{document}

\maketitle

\section*{SUMMARY}

Nonlinear pulse propagation in multimode fibers (MMFs) offers a compact, low-cost route to broadband, tunable femtosecond light, but most control schemes act by changing the spatial mode composition, typically resulting in irregular or speckled beams in exchange for maximal spectral tunability. Here we introduce a complementary mechanism: bending-induced local dispersion modification of a high-order mode (HOM) to steer the spectrum while keeping the spatial mode fixed. We launch an LP$_{0,7}$ mode into a step-index MMF and apply programmable macrobends near the input. With a standard Yb pump at 1030~nm, this yields spatially clean, continuous spectral tuning across 700–1350~nm, while the output profile remains Bessel-like and robust to reconfiguration of controlled bends. A perturbative model explains the observed spatial–spectral decorrelation, showing that moderate curvature produces first- and second-order shifts in group delay and group-velocity dispersion of the HOM with minimal change in its modal composition; these dispersion shifts control soliton fission, dispersive-wave emission, and the soliton self-frequency shift. We further validate application utility by driving multicolor, extended-depth-of-focus multiphoton microscopy directly from this all-fiber source. To our knowledge, this is the first demonstration of bending-induced dispersion modification, rather than mode mixing, used to tune MMF supercontinuum spectra without sacrificing beam quality, laying the foundation for an alternative pathway to tunable femtosecond illumination for imaging and spectroscopy.

\section*{KEYWORDS}


Multimode fiber, tunable supercontinuum, broadband Bessel beam, local dispersion engineering, multicolor volumetric nonlinear imaging

\section*{INTRODUCTION}

Nonlinear pulse propagation in optical multimode fibers (MMFs) has emerged as a compelling platform for controlling ultrashort pulses across the spectral, spatial, and temporal domains~\cite{Krupa2019,Wright2022}. MMFs support a vast number of spatial modes, enabling complex light--matter interactions and nonlinear dynamics, including soliton dynamics, intermodal four-wave mixing, and supercontinuum generation~\cite{Rishj2019,Wu2023,Demas2017,Perret2018,Wright2015,Eftekhar2017,Eslami2022}. These phenomena offer a low-cost and compact means to generate broadband, structured light fields essential for applications ranging from fiber lasers to biological imaging~\cite{Wright2017,Cao2023,Haig2023,Xu2013,Wang2014,Liu2017,Moussa2021}.

A strong interest exists in exploring the potential of MMFs for broadband spectral tunability, particularly for developing low-cost, high-power femtosecond sources for bioimaging~\cite{Xu2013,Wang2014,Liu2017,Moussa2021}. Multimodal nonlinear pulse propagation in MMFs allows wide spectral variations through alteration of the spatial mode composition, making MMFs a natural medium for this application. Leveraging this spatial control lever together with an array of nonlinear processes, prior works have demonstrated broadband manipulation using either spatial light modulators (SLMs) to shape the input wavefront~\cite{Wright2015,Tzang2018,Wei2020,Chen2023,Bender2023} or mechanical bending to induce local perturbations~\cite{Finkelstein2023,Qiu2024}. 
Both approaches harness the intrinsic space–time coupling of multimodal propagation described by the generalized multimode nonlinear Schrödinger equation~\cite{Poletti2008,Horak2012} (GMMNLSE), enabling rich, high-dimensional control over spatio--spectral--temporal output states. At the same time, since spectral changes are the direct consequence of varying the spatial profile, solely maximizing spectral tunability can lead to irregular or speckled beam profiles, and jointly optimizing beam quality and spectrum becomes a non-trivial high-dimensional search~\cite{Wei2020,Chen2023,Bender2023,Tegin2020}. These observations motivate a complementary pathway explored here—can we find an operating regime in which spectral steering is available while a stable, clean spatial mode can be maintained?

Here we identify and exploit a different control knob: when the launched field is dominated by a disorder-resilient high-order fiber mode (HOM), moderate macrobending primarily modifies the mode’s local dispersion characteristics—group delay and group velocity dispersion (GVD)—rather than its modal composition. In this regime, bending steers the nonlinear frequency conversion while the spatial profile remains essentially unchanged, decoupling spectral tuning from beam quality (Figure~\ref{fig:cascaded_control}a). We implement this idea in a step-index MMF (SI-MMF) pumped by an accessible Yb laser near 1030 nm, launching an LP$_{0,7}$ mode and placing programmable bends early along the fiber where nonlinear frequency conversion is most susceptible to perturbations. We obtain continuous, spatially clean tuning across 700–1350 nm with reproducible operation (Figure~\ref{fig:cascaded_control}b). We then demonstrate multicolor, extended-depth-of-focus (EDOF) nonlinear microscopy across standard near infrared windows directly from this fiber platform. 
Conceptually, the mechanism differs from prior bending-based methods in MMFs that intentionally scramble spatial modes~\cite{Finkelstein2023,Qiu2024}. By contrast, we operate where the spatial eigenvector is quasi-invariant and curvature perturbs the effective index landscape, producing shifts in the first- and second- dispersion that redirect soliton fission, dispersive-wave generation (DWG), and soliton self-frequency shift (SSFS)—thereby tuning the spectrum without speckle formation. Intuitively, moderate mechanical bending perturbs the waveguide geometry and breaks its symmetry, thereby modifying the waveguide dispersion and consequently influencing supercontinuum generation. At the same time, because the induced refractive-index perturbation varies slowly along the fiber, the process is largely adiabatic: the spatial profiles of disorder-resilient HOMs are mostly unchanged under bending, while their modal dispersion is altered.
Our perturbative treatment and experiments are consistent with classic analyses that bending alters propagation constants and dispersion in guided-wave optics, now observed to be a practical spectral control lever in the multimode, HOM-dominated regime.

In summary, here we show that moderate macrobending can act primarily through local dispersion modification of a HOM, enabling spectral tuning with a fixed spatial mode. We (i) experimentally demonstrate bending-controlled supercontinuum tuning in an SI-MMF while maintaining a uniform Bessel-like output from an LP$_{0,7}$ launch with a Yb pump, (ii) provide a perturbative model that explains the observed spatial--spectral decorrelation in the HOM regime, and (iii) validate utility by driving multicolor, EDOF multiphoton imaging directly from this all-fiber source. To our knowledge, this is the first use of bending-induced dispersion—rather than mode mixing—to steer MMF supercontinuum spectra without compromising beam quality.

\section*{RESULTS}

\subsection*{Observation of tunable MMF supercontinuum with a fixed HOM profile}

\begin{figure}[h!]
    \centering
    \includegraphics[width=0.90\linewidth]{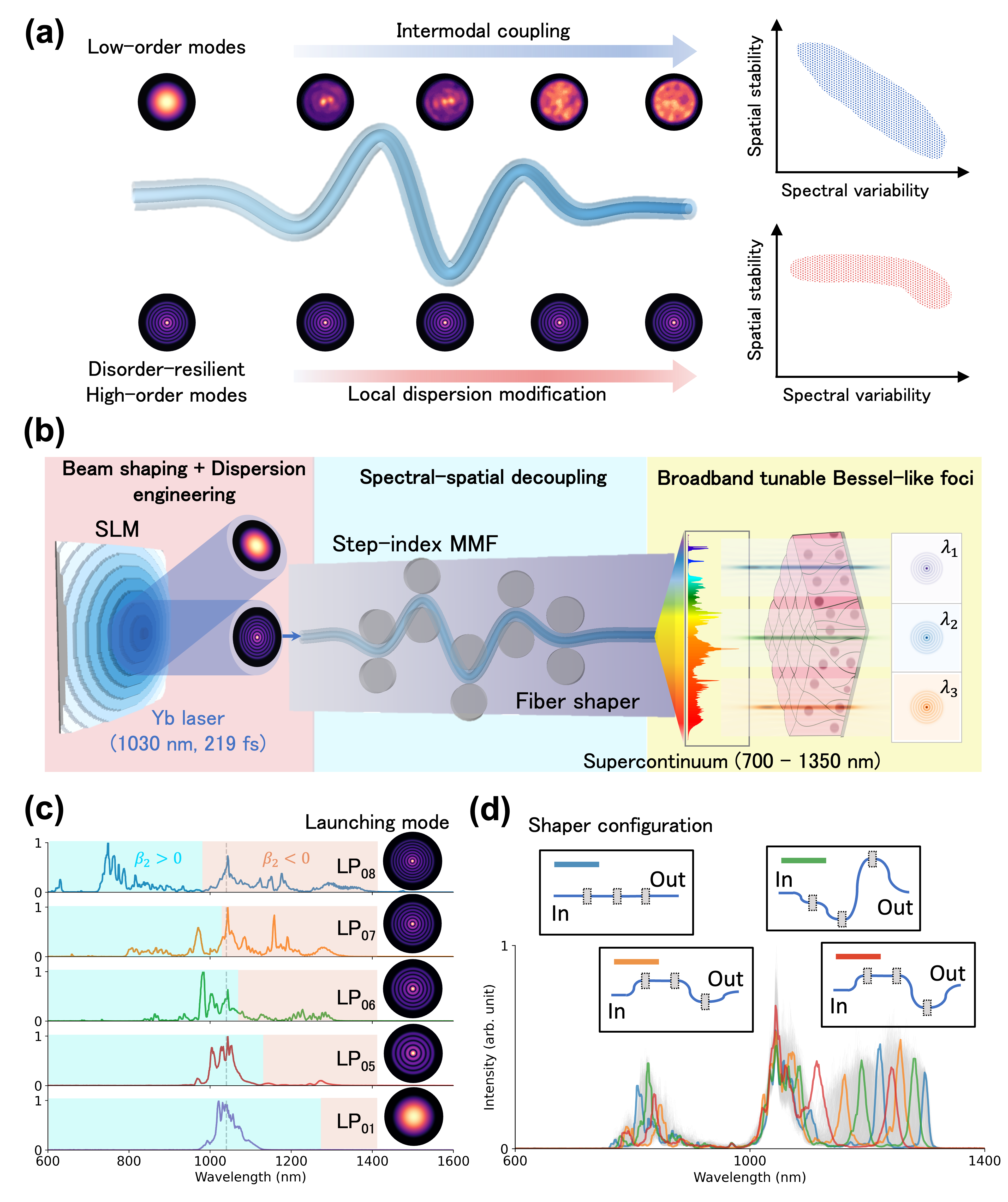}
    \caption{\textbf{Bending-controlled spectrally tunable HOM supercontinuum.}
    \textbf{(a)} Schematic of nonlinear pulse control of a HOM in MMF via local disperison modification.
    \textbf{(b)} Schematic of experimental setup for bending-controlled HOM supercontinuum. First, a liquid-crystal SLM transforms the pump pulse from a Yb laser into a high-order fiber mode in an SI-MMF. Then, multiple motorized fiber shapers apply local bending to modulate the spectrum without scrambling the spatial profile, yielding broadband tunable Bessel-like foci at the output. 
    \textbf{(c)} Dispersion engineering via launching high-order fiber modes for broadband supercontinuum evolution. The traces show experimental spectra in a 70-cm SI-MMF pumped with 400~nJ pulses. 
    \textbf{(d)} Tunable supercontinuum via mechanical perturbation. A 9-meter SI-MMF and a lower pump energy of 65~nJ were used to emphasize spectral contrast resulting from mechanical perturbation.}
    \label{fig:cascaded_control}
\end{figure}

This method originates from our experimental observation that the output supercontinuum is highly tunable under perturbation while maintaining a HOM profile in an SI-MMF. As illustrated in Figure~\ref{fig:cascaded_control}a, this phenomenon is in stark contrast to the behavior of the mixture of low-order modes in our prior work~\cite{Qiu2024}, which typically form distinct speckles at different wavelengths due to intermodal coupling and chromatic dispersion. 
We hypothesized that the underlying mechanism should differ from the existing understanding that mode mixing causes changes in the spectrum of the supercontinuum.

\begin{figure}[h!]
    \centering
    \includegraphics[width=0.95\linewidth]{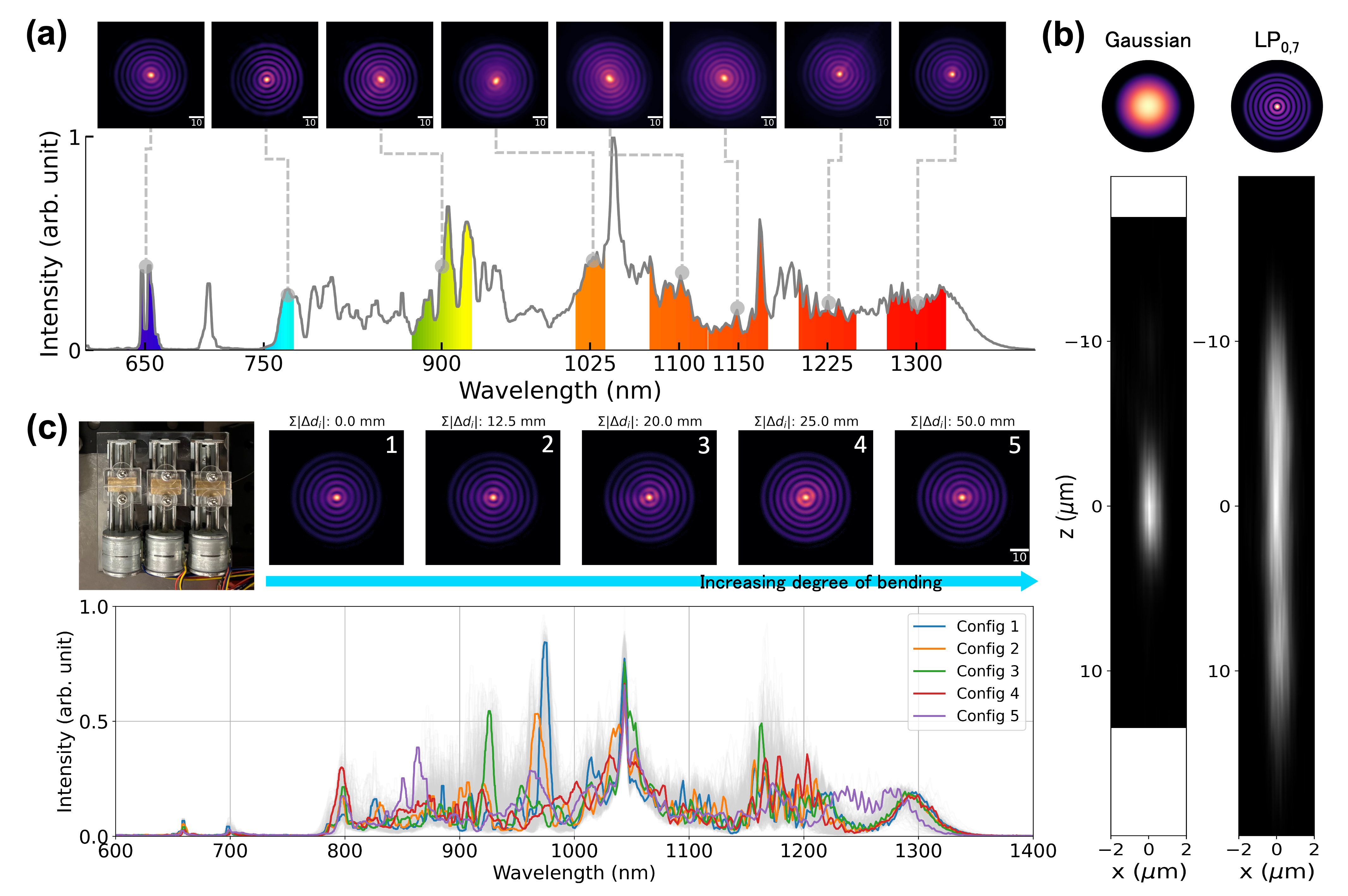}
    \caption{\textbf{Broadband tunable Bessel-like foci via disorder-resilient LP$_{0,7}$ mode pump.}
    \textbf{(a)} Broadband Bessel-like foci with an LP$_{0,7}$ mode pump in an SI-MMF. Beam profiles were measured under different bandpass filters. 
    \textbf{(b)} The two-photon PSFs at 1100~nm showing axial FWHM of 4.2~\textmu m for the Gaussian focus and 24~\textmu m for the Bessel-like focus. 
    \textbf{(c)} Moderate bending alters the spectrum while preserving the high-order spatial mode. Bending is quantified by the total displacement $\sum |\Delta d_i|$, where $d_i$ is the displacement of the i-th shaper with respect to the previous shaper motor. The spectra under 730 fiber shaper configurations are plotted in gray traces, with five highlighted spectra associated with the above spatial profiles. Inset (top-left): photograph of the precision laser cut motorized fiber shapers.}
    \label{fig:broadband_Bessel}
\end{figure}

The experimental setup is illustrated in Figure~\ref{fig:cascaded_control}b. We launched 219~fs, 400~nJ pulses at 1030~nm into a 70-cm silica SI-MMF with a 50-\textmu m core and a numerical aperture (NA) of 0.22. We introduced programmable mechanical perturbations along the fiber axis to modulate the output. Based on the previous setup, we incorporated two major changes. First, we improved the bending control using precision laser-cut motorized fiber shapers, a more compact and precise version of the previously 3D-printed design~\cite{Qiu2024} (the inset in Figure~\ref{fig:broadband_Bessel}c; see Methods~\nameref{method:fiber_shaper} for details), to apply controlled bends with radii down to 1~cm. We further enhanced the mechanical stability and reproducibility of the translational actuators with a post-stabilization unit, enabling precise and repeatable positioning as well as real-time adaptive adjustments. As shown in Supplementary Figure~S1a, the reproducibility test yields an averaged spectral cosine similarity exceeding 0.99 after 100 repeated configuration switches. The long-term spectral stability is also confirmed in Supplementary Figure~S1b, which shows minimal drift over a 3-hour continuous operation. Additionally, the mechanical layout was redesigned to position the shapers closer to the fiber input, within the first ten centimeters, where the nonlinear pulse evolution is most sensitive to perturbations.

The other key difference is the launching condition. We selected the LP$_{0,7}$ mode as the launching mode, by which the zero-dispersion wavelength (ZDW) shifts from $\sim$1300~nm (in the fundamental LP$_{0,1}$ mode) to $\sim$1030~nm (Figure~\ref{fig:cascaded_control}c)~\cite{Qiu2024}, enabling anomalous-dispersion pumping with a Yb laser. Once in the anomalous-dispersion regime, initial self-phase modulation seeds modulation instability and soliton fission~\cite{Dudley2006,Agrawal2013-CH13}, leading to rapid spectral broadening within the first few tens of centimeters. The resulting fundamental solitons undergo Raman-driven soliton self-frequency shift (SSFS)~\cite{Gordon1986}, continuously extending beyond 1350~nm, while phase-matched dispersive-wave emission fills the 700--900~nm band. Together, these processes yield a continuous supercontinuum spanning 700 to 1350~nm, establishing a broad spectral base for multicolor operation. 
Across this entire spectral range, the output maintains a uniform, ring-shaped profile (Figure~\ref{fig:broadband_Bessel}a), attributed to the high modal purity of the LP$_{0,7}$ mode achieved by the negative axicon method (Supplementary Note 2). The high mode purity suppresses intermodal coupling—unlike the speckled output patterns commonly observed in multimode fibers with unshaped input beams~\cite{Morales-Delgado2015}. As shown in Figure~\ref{fig:broadband_Bessel}b, two-photon point spread function (PSF) measurements at 1100~nm display an axial depth of focus of approximately 24~\textmu m, about six-fold longer than that of a Gaussian reference, while retaining submicron lateral resolution ($\sim$0.95~\textmu m full width at half maximum, FWHM). This characterization confirms that the space-time properties of the output beam resemble the non-diffracting Bessel-like behavior.

The stability of the HOM supercontinuum is quantified in Supplementary Figure~S1c-e in three conditions: (1) varying input pulse energy, (2) varying input beam pointing, and (3) varying time points. Among all, input beam pointing is the most critical parameter that can give rise to a considerable spectral variation with a 0.13 mrad deviation. It arises from the fact that an oblique input beam cannot excite a pure HOM due to the break of rotational symmetry. Consequently, a high-quality laser with less than 50 \textmu rad beam pointing fluctuation, a temperature management system for the SLM, and a shielded beam path are essential for stable operation. On the other hand, a small fluctuation in input pulse energy (e.g., $<$1\% for most of the commercial Yb lasers) does not yield a dramatic change, and the ensemble average of the spectrum shows a low spectral fluctuation (an averaged spectral cosine similarity of 0.985).

The spatial and spectral profiles under different fiber shaper configurations are shown in Figure~\ref{fig:broadband_Bessel}c. The high modal purity and the near-Bessel spatial profile are preserved when localized perturbations are applied by the fiber shapers, while the spectrum rapidly deviates from the initial profile. Unlike adaptive wavefront shaping, which can largely affect the output spatial profile and reduce coupling efficiency, the bending-controlled HOM profile retains more than 80\% intensity overlap with an ideal LP$_{0,7}$ mode under most configurations, with coupling efficiency fluctuations remaining below 8\%. Supplementary Figure~S2c quantifies the spectral deviation and spatial stability of the supercontinuum under various bending configurations, displaying clear evidence of spatial--spectral decorrelation when applying mechanical perturbation to an HOM, a unique operating regime that has not been explored.  

\subsection*{Spectral tuning via bending-induced local dispersion modification}

The observed tunable HOM supercontinuum exhibits stable spatial output while its spectral distribution is effectively tuned (Figure~\ref{fig:broadband_Bessel}c and Supplementary Figure~S2c), in contrast to previous studies on bent MMFs~\cite{Finkelstein2023,Qiu2024} where randomized mode composition leads to pronounced spectral variations.
These observations highlight a fundamentally different tuning mechanism: rather than an intentionally scrambled mode profile, in our operating regime, the spatial eigenmode remains quasi-invariant under bending because moderate curvature introduces a slowly varying refractive-index perturbation along the fiber, which can be viewed as effectively adiabatic. In contrast, the propagation constants (i.e., the eigenvalues of the waveguide operator) are modified through changes in the effective optical path. The resulting shifts in group velocity and GVD subsequently reshape the nonlinear dynamics involved in supercontinuum generation, including soliton fission, DWG, and SSFS.

Compared with low-order modes, HOMs are intrinsically more resilient to bending-induced disorder for several reasons. First, their larger propagation-constant separations reduce the maximum intermodal power transfer rate according to the coupled mode theory~\cite{Yariv1973}. Second, stronger modal dispersion among HOMs accelerates temporal walk-off, particularly in the femtosecond regime. Third, moderate bending can be approximated as a slowly varying spatial perturbation of the refractive index~\cite{Heiblum1975,Kiiveri2022}; for rapidly oscillating HOM fields, this perturbation largely cancels in the overlap integral, further suppressing intermodal coupling. As a result, intermodal coupling remains weak, and bending-induced dispersion modification becomes the dominant mechanism governing the nonlinear pulse evolution.
Numerical modeling based on the modified Helmholtz equation~\cite{PlschnerMartinandTyc2015,Popoff2024} (Supplementary Figure~S2a--b) confirms that after localized deformation, HOMs such as LP$_{0,7}$ recover their spatial profiles, whereas low-order modes tend to scatter irreversibly into adjacent modes. This modal analysis is consistent with the experimental characterization in Supplementary Figure~S2d, where the HOM profiles present greater robustness to bending across considerable spectral variations.

\begin{figure}[h!]
    \centering
    \includegraphics[width=0.95\linewidth]{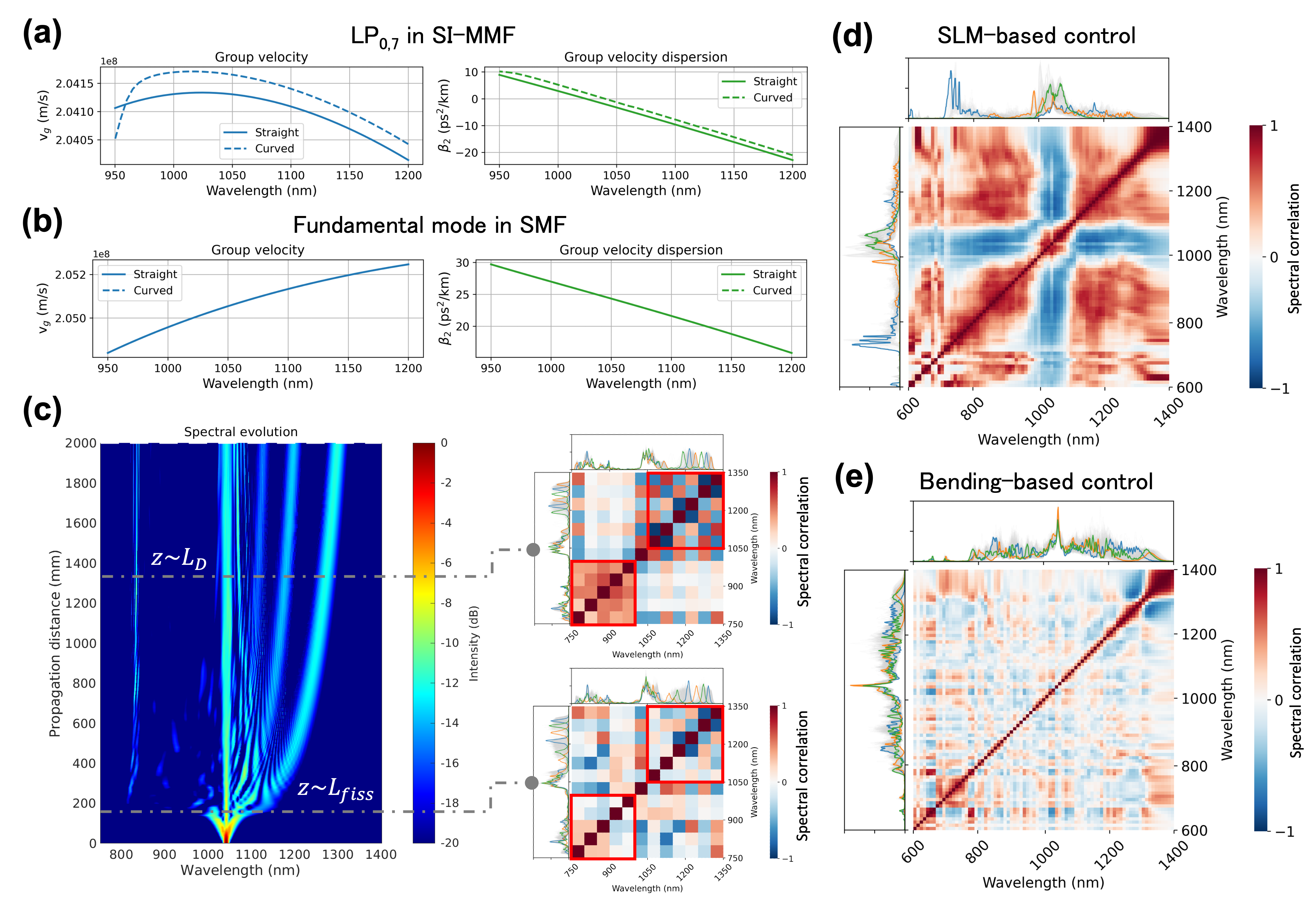}
    \caption{\textbf{Bending-induced dispersion modification}
    \textbf{(a-b)} The 1$^{st}$ and 2$^{nd}$ order dispersion curves of the LP$_{0,7}$ mode in a 50-\textmu m-core SI-MMF (a) and a 5-\textmu m-core SMF with 0.1 NA (b).
    \textbf{(c)} Spectral decorrelation via perturbation at different distances. Left: the simulated spectral evolution (pumped by the LP$_{0,7}$ mode) showing different timings of soliton fission/dispersive wave generation and SSFS. Right: the spectral correlation maps obtained by applying mechanical perturbation at soliton fission length ($L_{fiss}$) and dispersive characteristic length ($L_{D}$). Pearson correlation coefficients were computed with a 50~nm binning width, with a total of 730 spectra for each distance. The spectral evolution simulation was performed by a MATLAB GMMNLSE solver~\cite{Wright2018}.
    \textbf{(d--e)} Spectral correlation maps of the supercontinuum generated by SLM modulation (d) and shaper modulation (e). Pearson correlation coefficients were computed with a 10~nm binning width, with a total of 800 and 730 spectra were for the SLM- and shaper-modulated cases, respectively. }
    \label{fig:local_dispersion}
\end{figure}

Even in the absence of intermodal coupling, the new control platform demonstrated a wide spectral tuning range. Perturbative analysis (see Methods~\nameref{method:perturbative} and Supplementary Note 1) reveals that in weakly perturbed MMFs, changes in group velocity and GVD stem from the superposition of multiple participating eigenmodes during bending. This bending-induced spectral tuning through local dispersion modification is a unique property of MMFs. In SMFs, the absence of higher-order modes limits the induced dispersion change. This finding is consistent with prior reports showing that relatively strong bending is required to achieve even a fractional change in GVD in SMFs~\cite{Gil-Molina2018}. The dispersion simulations (Figure~\ref{fig:local_dispersion}a-b) further support this analysis, showing that macrobending with a radius $R = 1$~cm can induce a GVD change of $\Delta\beta_2 \sim 2~\mathrm{ps}^2/\mathrm{km}$ for the LP$_{0,7}$ mode in an MMF, while leaving the fundamental mode in an SMF largely unaffected. 

The effect of local dispersion modification can be further verified through positional bending on a longer fiber, where distinct nonlinear processes are spatially separated along the propagation axis. Figure~\ref{fig:local_dispersion}c shows the spectral correlation obtained when bending is applied at two locations in a 9-meter SI-MMF pumped with an LP$_{0,7}$ mode: (1) early bends near the soliton fission length ($L_{fiss} \sim L_{D}/N$, where $N$ is the soliton order) and (2) later bends near the dispersive characteristic length ($L_{D} = T_0^2 / |\beta_2|$, where $T_0$ is the pulse width and $\beta_2$ the GVD)~\cite{Dudley2006}. Upon fiber bending, wavelength pairs that respond concurrently exhibit a high correlation, with a positive values indicating shifts in the same direction and negative values indicating the opposite. A high correlation often implies that the wavelength pair is involved in the same nonlinear process, making direct energy transfer between them possible. The spectral correlation maps clearly distinguish the respective effects of these two regions on the supercontinuum evolution. Immediately after the onset of self-phase modulation and soliton formation (within the first tens of centimeters), dispersive waves are seeded by phase-matched solitons during soliton fission and rapidly stabilize~\cite{Dudley2006,Agrawal2013-CH13}. Early bending ($z\sim L_{fiss}$) modifies the local dispersion and consequently the phase-matching condition between solitons and dispersive waves, leading to strong tunability across both the red-shifted and blue-shifted bands. 
This behavior is also verified under a low-power pump, at which dispersive wave generation are much more pronounced than other nonlinear effects due to its shorter characteristic length. In Supplementary Figure~S3, wavelength shifts of the red-shifted solitons and the blue-shifted dispersive waves are clearly resolved under controlled bending, as a result of modified dispersion properties that alter the phase-matching condition.
In contrast, the SSFS accumulates gradually over tens to hundreds of centimeters, following an incremental rate proportional to $\mathrm{d}v/\mathrm{d}z \propto -|\beta_2|/T_0^4$~\cite{Gordon1986,Agrawal2013-CH5}. The dispersion modification induced by later bending ($z\sim L_{D}$) therefore predominantly affects the red-shifted solitons while leaving the dispersive waves largely unchanged. The constant soliton pulse duration in Supplementary Figure~S4 confirms that the spectral tunability is mainly attributable to dispersion changes in this regime. This position-dependent behavior not only corroborates the mechanism of bending-induced dispersion modification but also highlights the potential for orthogonal spectral tuning of individual supercontinuum components with distinct characteristic lengths.

Moreover, spectral tuning achieved through wavefront shaping and mechanical perturbation exhibits markedly different spectral correlation behaviors. Specifically, wavefront shaping induces global changes in dispersion and phase-matching conditions, shifting all spectral bands in a concerted, highly correlated manner (Figure~\ref{fig:local_dispersion}d). Notably, the structured long-range correlation spanning the entire spectral range aligns well with the spectra shown in Figure~\ref{fig:cascaded_control}c. The pump wavelength shows a strong negative correlation between both the solitons and dispersive waves, whereas the latter two show a strong positive correlation. This behavior arises from the global dispersion characteristics that govern soliton dynamics and the formation of phase-matched dispersive waves. In contrast, mechanical perturbation acts locally on the fiber, resulting in the absence of long-range spectral correlations (Figure~\ref{fig:local_dispersion}e) , suggesting that inter-wavelength energy transfer is no longer primarily governed by the same mechanism as in wavefront shaping; instead, energy redistribution among neighboring wavelengths is more prominent, resulting in substantially weaker dependencies between distant spectral components. Supplementary Figure~S5 further illustrates the decorrelation between solitons and dispersive waves, as well as among multiple solitons. Unlike wavefront shaping, where the solitons and dispersive wavelengths show strong linear correlation, mechanical perturbation allows access to a wide range of wavelength combinations, thereby offering a more flexible way to spectral tuning.

\subsection*{Optimization of broadband tunable supercontinuum for multicolor nonlinear imaging}

\begin{figure}[h!]
  \centering
  \includegraphics[width=0.95\textwidth]{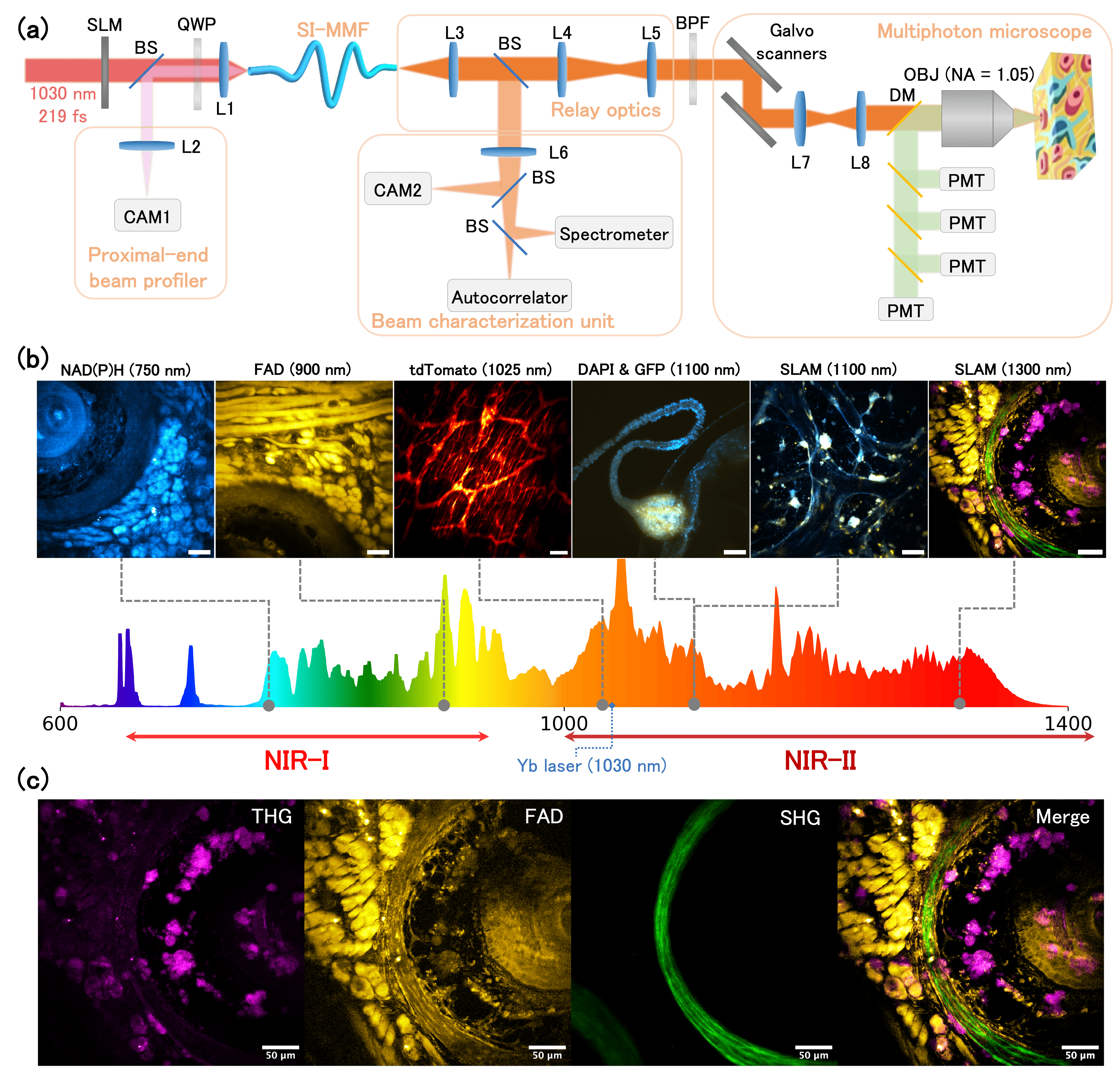}
  \caption{ \textbf{Versatile biological contrasts via broadband tunable femtosecond pulses.}
  \textbf{(a)} System schematic of nonlinear microscopic imaging using broadband Bessel-like foci. Proximal end: L1--L2 are lenses with a focal length of 19 and 200 mm, respectively. Relay optics: L3--L5 are lenses with a focal length of 10, 500, and 400 mm, respectively. Beam characterization unit: L6 is a lens with a focal length of 300 mm. Multiphoton microscope: L7--L8 are lenses with a focal length of 30 and 100 mm, respectively. BS: beam splitter. QWP: quarter-wave plate. CAM: camera. BPF: bandpass filter. DM: dichroic mirror. PMT: photomultiplier tube. OBJ: objective lens.
  \textbf{(b)} Nonlinear microscopy modalities enabled by the broadband Bessel-like foci. Scale bars: 50~\textmu m.
  \textbf{(c)} SLAM imaging of fixed mouse whisker pad tissue using 1300~nm excitation. The emission filters are 430~$\pm$~5~nm (THG), 530~$\pm$~27.5~nm (FAD), and 650~$\pm$~20~nm (SHG). The pixel size is 500~nm, and the dwell time is 400~\textmu s.
  }  
  \label{fig:multicolor_imaging}
\end{figure}

The broadband Bessel-like foci spanning 700 to 1350~nm enable access to multiple key biological contrasts in label-free nonlinear microscopy. For example, dispersive waves can probe the two-photon fluorescence (2PF) of metabolic coenzymes, NAD(P)H at 750~nm and FAD at 900~nm, while solitons in the 1100--1300~nm range align with SLAM (simultaneous label-free autofluorescence multiharmonic) imaging~\cite{You2018,Liu2024}. We employed these broadband Bessel-like foci in our home-built multiphoton microscope, as shown in the system schematic in Figure~\ref{fig:multicolor_imaging}a. The corresponding images of various modalities are displayed in Figure~\ref{fig:multicolor_imaging}b. A relay system was incorporated to match the beam size of the Bessel-like foci to the back aperture of the objective, ensuring optimal signal generation efficiency and resolution. A pulse characterization unit—comprising a spectrometer, a beam profiling camera, and an autocorrelator—was used to characterize the output pulses and provide real-time feedback.

We leveraged the spectral tunability enabled by the fiber shapers to optimize the spectral density of target wavelengths for improved imaging performance. To spectrally enhance a desired band (e.g., 1300~$\pm$~25~nm), we formulated the task as an adaptive optimization problem with real-time feedback across three fiber shapers (with the SLM phase mask fixed). The pulse characterization unit was used to measure the second-harnomic generation (SHG) signal in a beta-barium borate (BBO) crystal as the figure of merit (see Methods~\nameref{method:adaptive_opt}), which jointly optimizes spectral density and pulse duration for nonlinear microscopy. We employed the Covariance Matrix Adaptation–Evolution Strategy (CMA-ES)~\cite{Hansen2019}—a gradient-free, population-based optimizer that iteratively adapts a multivariate Gaussian search distribution—to explore the actuator space. Compared to exhaustive or greedy search strategies, CMA-ES demonstrates robust performance in handling noisy measurements and optimizing complex, nonconvex objective functions. At 1300~nm, three fiber shapers offer an 18-fold spectral tunability, defined as the ratio between maximal and minimal spectral density, and CMA-ES reliably finds configurations that outperform the unshaped baseline (Supplementary Figure~S6b--c). Pulses with energies up to 20~nJ and durations of approximately 100~fs can be delivered to the back aperture (Supplementary Figure~S4). The resulting multimodal images of a fixed mouse whisker pad tissue (Figure~\ref{fig:multicolor_imaging}c) simultaneously display third-harmonic generation (THG, magenta), three-photon fluorescence from FAD (3PF-FAD, yellow), and SHG (green) with high contrast and submicron resolution. The near–transform-limited pulse duration is attributed to the self-regulating nature of solitons, which maintain constant pulse width by balancing dispersion with nonlinear phase shifts.

For this imaging task, we used the SHG signal strength as the figure of merit. Alternatively, various objective functions can be incorporated to explore the high dimensionality of the tunable source. As a proof of concept, we demonstrated a multi-band joint optimization strategy as a potential solution for multicolor or multiplexed excitation (Supplementary Figure~S6d--g). Specifically, we selected the logarithmic sum of spectral intensities in Equation~\ref{eqn:fom} (see Methods~\nameref{method:adaptive_opt}) across three target bands—900, 1100, and 1325~nm. Using this criterion, a 6.5-fold overall spectral enhancement was achieved with the fiber shapers, compared to the SLM-only configuration. These improvements are primarily attributable to the decorrelation between solitons and dispersive waves, reducing the trade-offs between individual objectives. Spectral tunability analysis (Supplementary Figure~S7) further confirms that the integration of wavefront shaping with mechanical perturbation yields a larger dynamic tuning range. In practice, the CMA-ES algorithm typically converges within 40 to 60 iterations, offering a speedup of 100 to 1000 times compared to an exhaustive search, which would otherwise require tens to hundreds of thousands of measurements.

\subsection*{Broadband tunable Bessel-like foci enable multicolor volumetric nonlinear microscopy}

\begin{figure}[h!]
  \centering
  \includegraphics[width=0.90\textwidth]{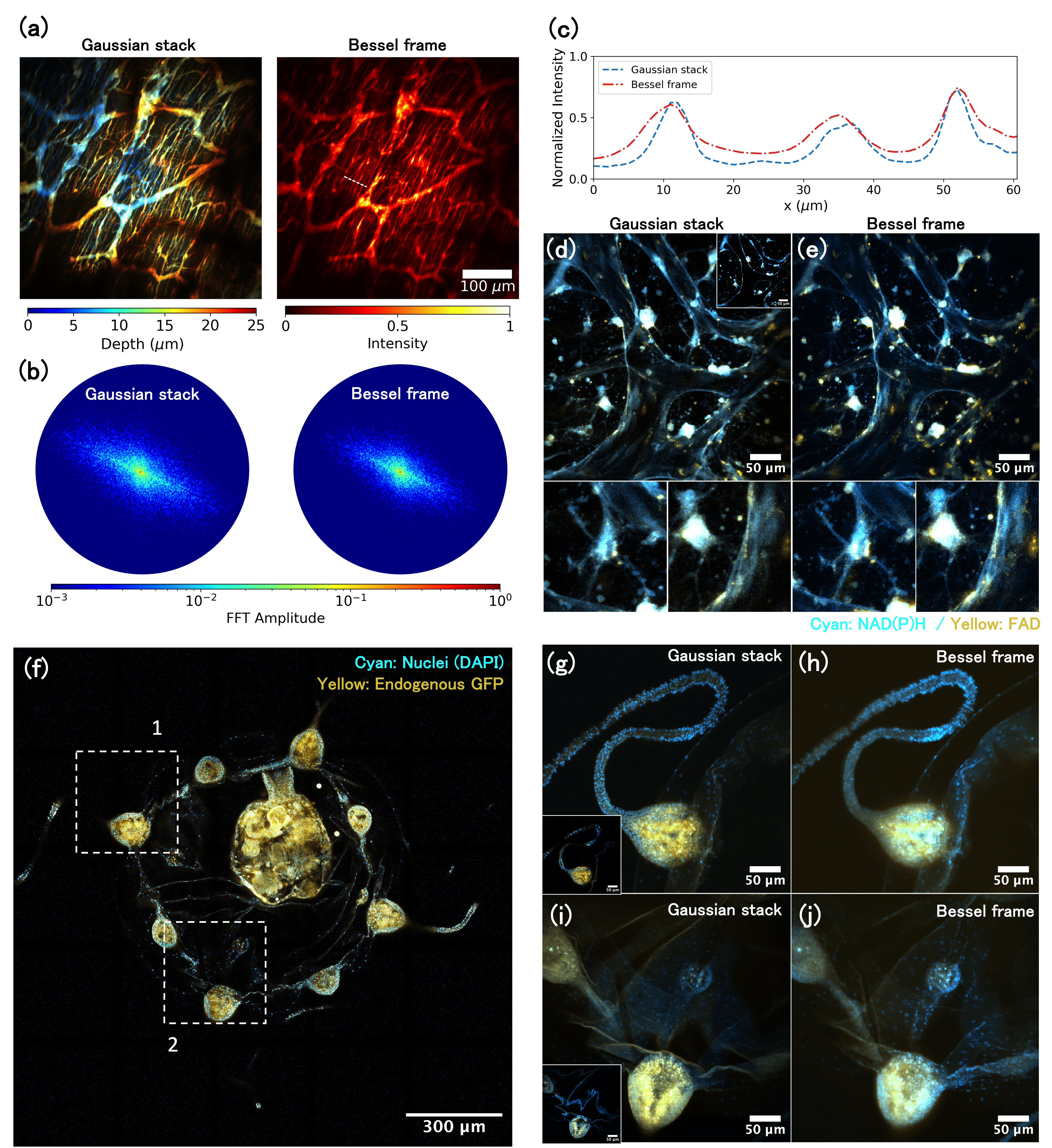}
  \caption{ \textbf{Volumetric nonlinear imaging via broadband Bessel-like foci.}
  \textbf{(a--c)} Two-photon volumetric imaging of an enteric nervous system (tdTomato-expressed ganglion network) using 1025~nm excitation. The image volume is 500~$\times$~500~$\times$~25~\textmu m$^3$ with a 500~nm lateral pixel size. The $z$ step size is 1~\textmu m for the Gaussian stack. \textbf{(b)} Fourier analysis of the images in (a). \textbf{(c)} Line profiles of the white-dashed cross sections in (a). 
  \textbf{(d--e)} Metabolic imaging of a blood-brain-barrier model using 1100~nm excitation. The image volume is 300~$\times$~300~$\times$~25~\textmu m$^3$ with a 500~nm lateral pixel size. The $z$ step size is 1~\textmu m for the Gaussian stack. The inset in (d) shows a single $z$-frame. The zoomed-in views highlight astrocytes and endothelial cells in the field of view. 
  \textbf{(f--j)} Multicolor imaging in a jellyfish nervous system using 1100~nm excitation. (f) shows the full jellyfish sample with a single $z$-frame. (g--h) highlight region 1, and (i--j) highlight region 2. The image volume is 300~$\times$~300~$\times$~30~\textmu m$^3$ with a 500~nm lateral pixel size and a $z$ step size of 1~\textmu m. The insets in (g) and (i) show single $z$-frames of the corresponding regions.
  }
  \label{fig:volumetric_imaging}
\end{figure}

After evaluating its capability for multicolor imaging, we investigated the performance of volumetric illumination in 3D biological samples. In Figure~\ref{fig:volumetric_imaging}a, we compare volumetric two-photon imaging at 1025~nm using the Bessel-like foci with conventional Gaussian 3D scanning in an enteric nervous system. The tdTomato-expressing ganglion network is clearly resolved in the Bessel projection, corresponding to an imaging volume of approximately 25~µm in depth, matching that of the Gaussian image stack. A Fourier analysis (Figure~\ref{fig:volumetric_imaging}b) and a cross-sectional line profile (Figure~\ref{fig:volumetric_imaging}c) confirm comparable lateral resolution between the Bessel-like and Gaussian beams, consistent with the PSF measurements in Figure~\ref{fig:broadband_Bessel}b.

We next examined volumetric metabolic imaging in a blood–brain-barrier model using 1100~nm excitation—an optimized wavelength for the simultaneous excitation of 3PF from NAD(P)H and 2PF from FAD. The 3D vasculature in this in vitro cell culture spans tens of micrometers axially, necessitating the volume information to study its microphysiology. As shown in Figure~\ref{fig:volumetric_imaging}d--e, the Bessel-like foci enable a 2.5D projection of vascular structures and various cell types, including endothelial cells, astrocytes, and pericytes, which are not fully captured by a 2D Gaussian scan (inset in Figure~\ref{fig:volumetric_imaging}d). Subcellular features, such as astrocyte endfeet and endothelial linings, are clearly visible in the highlighted regions.

Additionally, we demonstrated multicolor volumetric imaging at 1100~nm in a jellyfish nervous system, where neural and organ structures are distributed across a 3D volume up to several hundred micrometers thick (Figure~\ref{fig:volumetric_imaging}f). As illustrated in Figure~\ref{fig:volumetric_imaging}g--j, the Bessel-like foci capture distinct anatomical features within a single $\sim$30~µm axial range, including the tentacle (a whip-like structure with dense surface nuclei), the tentacle bulb (exhibiting endogenous green fluorescent protein in addition to surface nuclei), and the translucent bell (with sparse nuclei on the exterior membrane and denser nuclei on the interior membrane). Altogether, these demonstrations establish our broadband Bessel-like beam source as a versatile platform for multicolor volumetric microscopy, providing 2.5D projection imaging with submicron resolution and multiplexed contrast across a wide range of biological contexts.

\section*{DISCUSSION}

In this work, we demonstrated a bending-controlled tunable HOM supercontinuum, offering an alternative pathway to harness the high dimensionality in an SI-MMF for advanced light manipulation. The decoupled spatial and spectral properties arise from a new operating regime where the dispersion characteristics of a disorder-resilient HOM are modified through moderate macrobending. This approach allows for generation of broadband-tunable structured light, such as quasi-Bessel beams with an LP$_{0,m}$ mode , in an all-fiber configuration. Compared to conventional free-space Bessel beam generation methods using SLMs or axicons, the MMF-based approach launches the broadband Bessel-like foci at the pump wavelength, thereby circumventing limitations imposed by chromatic dispersion. Beyond Bessel beams, this strategy is generalizable to other structured beams. For instance, non-zero angular momentum modes (i.e., LP$_{n,m}$ modes with $n \neq 0$) can approximate vortex beams while simultaneously offering dispersion engineering to support broadband spectral evolution.

We employed an adaptive optimization scheme to maximize the SHG signal as a proxy for optimal multiphoton image quality. This objective effectively achieves joint optimization of both spectral and temporal properties tailored to second-order nonlinear processes. We further demonstrated multi-band joint optimization, showing enhancement across multiple spectral bands due to the spectral decorrelation afforded by the fiber shapers. The flexibility of the platform allows for alternative feedback metrics to be integrated, enabling full exploration of its high-dimensional control space. While we intentionally limited the number of fiber shapers to three to accelerate optimization convergence, increasing their number allows for more precise control over fiber geometry, thereby enabling more complex multi-objective optimizations to simultaneously tailor multiple pulse attributes. Although not the primary focus of this study, we anticipate that this approach, with appropriately designed objective functions, can be broadly applied to a multitude of scenarios demanding simultaneous spectral, spatial, and temporal customization of ultrashort pulses~\cite{Wei2020,Bender2023,Chen2023}.

Whereas we demonstrated a near-700~nm bandwidth using a Yb laser pump, this approach is not fundamentally limited to this range. In principle, solitons undergo continuous red-shifting, and longer wavelengths are achievable with sufficient fiber length. Interestingly, solitons in HOM profiles cease red-shifting beyond approximately 1400~nm due to the presence of a second, longer ZDW. SSFS terminates when a soliton approaches this second ZDW, regardless of increases in pump power or fiber length. Instead, Cherenkov radiation generates dispersive waves on the opposite side of the ZDW, which are rapidly attenuated due to the waveguide cutoff~\cite{vanHowe2007}. This limitation can be overcome by employing a high-NA multimode fiber, which alleviates waveguide cutoff via a larger core-cladding index contrast, or a custom double-cladding fiber engineered to support modes with a longer ZDW and cutoff wavelength.

Currently, the volumetric imaging rate using tunable Bessel-like foci remains limited (greater than 10~s per 300~$\times$~300~$\times$~30~\textmu m$^3$) due to the restricted power capacity and the absence of dispersion compensation in the objective lens, resulting in only marginal improvements in acquisition speed. Compared with Gaussian beams, the sidelobes of the Bessel-like foci reduce multiphoton generation efficiency. Consequently, achieving an equivalent signal-to-noise ratio typically requires either higher excitation power or a longer dwell time. In practice, we observed that a 2--2.5-fold higher power is needed to obtain a comparable two-photon signal yield. The maximum power throughput of the Bessel-like foci is ultimately constrained by the material damage threshold, which caps the input pulse energy below 400~nJ in our fiber. As shown in Supplementary Figure~S8, a high input pulse energy beyond the threshold will cause irreversible in-fiber damage due to self-focusing-induced defects~\cite{Ferraro2021}, validated by the photoluminescent spots along the fiber axis and the absence of damage-induced ablation feature at the fiber input. While the estimated peak intensity of the LP$_{0,7}$ mode at 700 nJ ($<$0.01 GW/cm$^2$ ) remains two orders of magnitude lower than the multiphoton ionization threshold~\cite{Cho1998}, the HOM supercontinuum shows considerably higher instability and even a reduced spectral bandwidth. One feasible solution is to increase the fiber core size, as the damage threshold scales proportionally with mode area. For instance, launching an LP$_{0,13}$ mode in the same multimode fiber with a doubled core diameter can shift the ZDW below 1030~nm while increasing power capacity approximately four-fold. Alternatively, employing shorter input pulses can enhance wavelength conversion efficiency by seeding higher-power solitons, given that soliton power is inversely proportional to the square of the pulse width.

\newpage

\section*{METHODS}

\subsection*{Experimental setup}
\phantomsection
\label{method:exp_setup}

The full experimental setup is illustrated in Figure~\ref{fig:multicolor_imaging}a. A 400~nJ pulse with a 219~fs duration generated by a Yb laser (CARBIDE-CB3, Light Conversion) served as the pump source. The spatial beam profile was shaped by an liquid-crystal SLM (Pluto-2.1, HOLOEYE) to generate a Bessel–Gaussian beam, which was projected onto a 70-cm SI-MMF (FG050LGA, Thorlabs) using a 19-mm doublet lens (AC127-019-B, Thorlabs). A quarter-wave plate was incorporated to increase the fiber damage threshold with a circularly-polarized pump. The fiber shaper device, equipped with three motorized actuators, was positioned 10~cm from the fiber input to introduce early bends for maximal spectral tunability. The output pulse was collected by a 10-mm achromatic lens, followed by a 4-f relay optics system with a 0.8$\times$ magnification to match the entrance pupil size of our home-built multiphoton microscope. Bandpass filters were placed after the fiber output to select wavelengths of interest, including 750~$\pm$~25~nm, 900~$\pm$~25~nm, 1025~$\pm$~12.5~nm, 1100~$\pm$~25~nm, and 1300~$\pm$~25~nm.

The pulse characterization unit shown in Figure~\ref{fig:multicolor_imaging}a consisted of two spectrometers (NIRQuest, Ocean Insight, 900--1700~nm; S2000, Ocean Insight, 500--1100~nm), a beam profiler (MakoG-040B, Allied Vision), and an autocorrelator (GECO, Light Conversion). The spectral responses of both spectrometers were precalibrated to compensate for scaling differences, and their data were stitched together at 1000~nm. Custom Python code was developed to control the SLM, fiber shapers, and pulse characterization devices, enabling adaptive monitoring and optimization of pulse properties.

All microscopic images reported in this work, including PSF characterizations, were acquired using a 1.05-NA water-immersion objective lens (XLPLN25XWMP2, Olympus). For SLAM imaging at 1300~nm excitation, three emission filters (430~$\pm$~5~nm, 530~$\pm$~27.5~nm, and 650~$\pm$~20~nm) captured THG, 3PF of FAD, and SHG signals, respectively. For SLAM imaging at 1100~nm excitation (including blood–brain barrier models), four emission filters (365~$\pm$~12.5~nm, 447~$\pm$~30~nm, 550~$\pm$~12.5~nm, and 609~$\pm$~27.5~nm) were used to collect THG, 3PF of NAD(P)H, SHG, and 2PF of FAD signals, respectively. For imaging the jellyfish nervous system, two filters (447~$\pm$~30~nm and 530~$\pm$~27.5~nm) were employed to capture DAPI and endogenous green fluorescent protein signals. The PSF measurement shown in Figure~\ref{fig:broadband_Bessel}b was acquired by imaging 100-nm green fluorescent beads.

To achieve high-purity LP$_{0,7}$ mode launching, we employed the negative axicon method, which theoretically introduces no spherical aberration—unlike the conventional binary phase plate (BPP) method~\cite{Brzobohat2008,Demas2015-HoM} (Supplementary Note 2). The SLM phase mask consisted of an axicon phase profile with an inverse linear phase ramp (i.e., a negative axicon), combined with a Fresnel lens mask. This configuration enables a precise 4$f$ relay to demagnify the Bessel–Gaussian beam, closely matching the LP$_{0,7}$ mode profile. Supplementary Figure~S9 presents the difference between the negative axicon method and the BPP method, highlighting the better mode matching performance of the former. In practice, we selected a magnification factor of 0.017 to achieve optimal mode matching. Notably, even slight misalignments, such as tilting or lateral displacement, can greatly degrade mode purity. To mitigate this, we iteratively refined the SLM parameters to maximize output mode purity. Under optimized launching conditions, we achieved a coupling efficiency of 80\% and a modal intensity overlap exceeding 93\% in a relaxed, unperturbed fiber.

\subsection*{Design of fiber shapers}
\phantomsection
\label{method:fiber_shaper}

The fiber shaper device used in this study was adapted from our previous work~\cite{Qiu2024}. The system comprises three linear lead-screw translational actuators (Hilitand, 2-phase, 4-wire), controlled by a microcontroller unit (Arduino Uno Rev3). To enhance mechanical stability and reproducibility, a post-stabilization unit was incorporated, which also minimized the risk of introducing sharp bends. The reproducibility evaluation was performed in Supplementary Figure~S1a, showing an averaged spectral cosine similarity above 0.99 and a standard deviation below 0.0013. The platform and fiber clamping components were fabricated from 5~mm and 3~mm acrylic sheets, respectively, using a high-precision laser cutter (Universal, PLS6.75). Control software was developed using the Arduino IDE and integrated with adaptive optimization algorithms written in Python, enabling automated operation and real-time adjustment. Compared to our prior implementation, the number of fiber shapers was reduced from five to three to accelerate optimization convergence. Additionally, the mechanical interface at the connecting dock was redesigned via laser cutting to enable closer placement to the fiber input facet, thereby enhancing spectral tuning efficiency.

\subsection*{Bending-induced dispersion modification - perturbative analysis}
\phantomsection
\label{method:perturbative}

To study the effect of bending on dispersion characteristics, we performed a perturbative analysis, with a full derivation provided in Supplementary Note 1. A brief summary is presented below. 

The eigenmode analysis for a curved fiber can be formulated as a generalized eigenvalue problem~\cite{PlschnerMartinandTyc2015,Popoff2024}:
\begin{equation}
    \label{eqn:eigenmode_problem}
    \hat{K} \textbf{c} = \lambda \hat{M} \textbf{c},
\end{equation}
where the Hermitian matrix $\hat{K}$ accounts for the geometrical modification due to curvature, and the diagonal matrix $\hat{M}$ represents the waveguide dispersion in the straight fiber. An eigenvector $\textbf{c}$ corresponds to a curved fiber mode expressed in the basis of the straight fiber modes, and the eigenvalue $\lambda = 1/\beta'^2$ is related to the propagation constant $\beta'$ of the perturbed mode.

The dispersion effect at a frequency $\omega_0$ can be treated as a small perturbation $\epsilon = \omega - \omega_0$ applied to all quantities in Equation~\ref{eqn:eigenmode_problem}. This leads to the following perturbative expansions:
\begin{subequations}
    \label{eqn:perturbation_epsilon}
    \begin{align}
        \hat{K}_\epsilon &= \hat{K} + \epsilon \hat{K}^{(1)} + \epsilon^2 \hat{K}^{(2)} + \mathcal{O}(\epsilon^3), \\
        \hat{M}_\epsilon &= \hat{M} + \epsilon \hat{M}^{(1)} + \epsilon^2 \hat{M}^{(2)} + \mathcal{O}(\epsilon^3), \\
        \lambda_\epsilon &= \lambda + \epsilon \lambda^{(1)} + \epsilon^2 \lambda^{(2)} + \mathcal{O}(\epsilon^3), \\
        \textbf{c}_\epsilon &= \textbf{c} + \epsilon \textbf{c}^{(1)} + \epsilon^2 \textbf{c}^{(2)} + \mathcal{O}(\epsilon^3),
    \end{align}
\end{subequations}
Solving for the first- and second-order corrections to the eigenvalue yields:
\begin{subequations}
    \label{eqn:perturbative_terms}
    \begin{align}
        \lambda^{(1)} &= \textbf{c}^{\dagger} (\hat{K}^{(1)} - \lambda \hat{M}^{(1)}) \textbf{c}, \\
        \lambda^{(2)} &= \textbf{c}^{\dagger} (\hat{K}^{(2)} - \lambda^{(1)} \hat{M}^{(1)} - \lambda \hat{M}^{(2)} ) \textbf{c} + \textbf{c}^{\dagger} ( \hat{K}^{(1)} - \lambda \hat{M}^{(1)} - \lambda^{(1)} \hat{M} ) \textbf{c}^{(1)},
    \end{align}
\end{subequations}
The dispersion coefficients are functions of $\lambda^{(1)}$ and $\lambda^{(2)}$:
\begin{subequations}
    \label{eqn:dispersion_coeff}
    \begin{align}
        \frac{d \beta'}{d \omega} &= -\frac{\beta'^3}{2} \lambda^{(1)}, \\
        \frac{d^2 \beta'}{d \omega^2} &= -\beta'^{3} \left( \lambda^{(2)} -\frac{3}{4} \beta'^2 \left(\lambda^{(1)}\right)^2 \right),
    \end{align}
\end{subequations}
The expressions in Equation~\ref{eqn:perturbative_terms} can be further simplified under the assumption that the dispersion of the eigenmode profiles in a weakly perturbed fiber is negligible compared to that of the propagation constants, i.e., $\hat{K}^{(1)} = \hat{K}^{(2)} = 0$ and $\mathbf{c}^{(1)} = 0$. This assumption is valid because the fiber mode profiles remain nearly unchanged over a relatively broad bandwidth, whereas the propagation constants vary more significantly. On the other hand, $\hat{M}^{(1)}$ and $\hat{M}^{(2)}$ are fully determined by the dispersion characteristics of the straight fiber:
\begin{subequations}
    \label{eqn:M_derivative}
    \begin{align}
        M_{ij}^{(1)} &= 2\beta_i \frac{d \beta_i}{d \omega} \delta_{ij}, \\
        M_{ij}^{(2)} &= \left[ \left( \frac{d \beta_i}{d \omega} \right)^2 + \beta_i \frac{d^2 \beta_i}{d \omega^2} \right] \delta_{ij},
    \end{align}
\end{subequations}
As a result, Equation~\ref{eqn:perturbative_terms} simplifies to:
\begin{subequations}
    \label{eqn:perturbative_terms_simplified}
    \begin{align}
        \lambda^{(1)} &= -2\lambda \sum_{i=1}^{N} {\beta_i \frac{d \beta_i}{d \omega} |c_i|^2}, \\
        \lambda^{(2)} &= -\sum_{i=1}^{N} \left[ 2\lambda^{(1)}\beta_i \frac{d \beta_i}{d \omega} + \lambda \left( \left( \frac{d \beta_i}{d \omega} \right)^2 + \beta_i \frac{d^2 \beta_i}{d \omega^2} \right) \right] |c_i|^2,
    \end{align}
\end{subequations}
Equation~\ref{eqn:perturbative_terms_simplified} delineates the bending-induced dispersion modification in MMFs, where the changes arise from the collective contributions of all participating modes (i.e., $c_i \neq 0$). In contrast, for SMFs where $N = 1$, this effect is absent.

In practice, obtaining the broadband dispersion characteristics of a straight SI-MMF is computationally inexpensive. Consequently, the above expansion treatment can be performed at each frequency bin, introducing a numerical error that scales with the cube of the frequency discretization.

\subsection*{Adaptive pulse optimization}
\phantomsection
\label{method:adaptive_opt}

To optimize the pulses for enhanced multiphoton signal yields, we used the SHG signal generated in a BBO crystal as the figure of merit. A photodiode recorded the signal strength, which was then passed to a CMA-ES algorithm to guide the control of the SLM and/or the fiber shapers. For each measurement, 10,000 pulses were acquired and averaged to suppress transient fluctuations in the system and stabilize the algorithm's convergence. Following Ref.~\cite{Hansen2019}, we selected a population size of 8 per iteration to ensure sufficient sampling diversity. For the SLM phase mask, the axicon apex angle and Fresnel lens focal length were tuned to vary the mode-launching conditions across LP$_{0,1}$ to LP$_{0,9}$. For the fiber shaper, three actuators were optimized, each with a travel range of $\pm$1~cm to prevent excessive fiber strain from bending.

To optimize multi-band spectral energy, as demonstrated in the Supplementary Figure~S6d--g, we integrated the spectral density of each band from the measured spectrum $S(\lambda)$ and calculated their logarithmic sum as the figure of merit:

\begin{equation}
\label{eqn:fom}
\mathrm{FoM} = \sum_{i=1}^{N} \log \left( \int_{\lambda_i} S(\lambda) \, d\lambda \right)
\end{equation}
where $\lambda_i$ denotes the $i$-th spectral band of interest. The logarithmic sum discourages the algorithm from favoring a single dominant band and instead promotes balanced optimization across all target wavelengths.

\newpage


\section*{RESOURCE AVAILABILITY}


\subsection*{Lead contact}


Requests for further information and resources should be directed to and will be fulfilled by the lead contact, Sixian You (sixian@mit.edu).

\subsection*{Materials availability}


This study did not generate new materials.

\subsection*{Data and code availability}


The data that support the findings of this study and the custom code used for analysis are available from the lead contact upon reasonable request.

\section*{ACKNOWLEDGMENTS}


This work was supported by NSF CAREER Award (2339338), and the CZI Dynamic Imaging program via the Chan Zuckerberg Donor Advised Fund (DAF) through the Silicon Valley Community Foundation (SVCF). Additional support was provided by NIA R01AG66768, R21AG072107, the Diacomp Foundation (Pilot Award, Augusta University), and a Pilot Grant from the Harvard Digestive Disease Core (SK). L.Y. acknowledges support from the Claude E. Shannon Award and the MathWorks Fellowship. H.C. and K.L. acknowledge support from the MathWorks Fellowship.

We express our gratitude to Rodger D. Kamm and his lab for providing the blood-brain-barrier models used in our imaging experiments.

\section*{AUTHOR CONTRIBUTIONS}


Conceptualization, L.-Y.Y. and S.Y.; 
methodology, L.-Y.Y., H.C., and S.Y.; 
investigation, L.-Y.Y., H.C., and K.L.; 
validation, L.-Y.Y., H.C., and C.L.; 
formal analysis, L.-Y.Y.; 
software, L.-Y.Y. and H.C.; 
resources, B.W., S.K., and S.Y.; 
writing--original draft, L.-Y.Y. and S.Y.; 
writing--review \& editing, L.-Y.Y. and S.Y.; 
visualization, L.-Y.Y.; 
funding acquisition, S.Y.; 
supervision, S.Y.


\section*{DECLARATION OF INTERESTS}


The authors declare no competing interests.



\section*{SUPPLEMENTAL INFORMATION INDEX}




\begin{description}
  \item Supplementary Notes 1-2 in a PDF
  \item Supplementary Figures S1-S9 and their legends in a PDF
\end{description}

\newpage

\bibliography{bibliography}

\newpage

\clearpage
\appendix
\pagenumbering{arabic} 
\renewcommand{\thefigure}{S\arabic{figure}}
\renewcommand{\thetable}{S\arabic{table}}
\renewcommand{\theequation}{S\arabic{equation}}
\setcounter{figure}{0}
\setcounter{table}{0}
\setcounter{equation}{0}

\input{supplementary} 

\end{document}

%% file: supplementary.tex
%
\section*{Supplementary Note 1: Dispersion analysis in a curved fiber via perturbation theory}
\subsection*{Modified Helmholtz Equation in a curved fiber}

Considering a curved fiber with a radius of curvature $\rho$, the scalar Helmholtz equation can be rewritten as~\cite{PlschnerMartinandTyc2015,Popoff2024}:

\begin{equation}
    \label{eqn:modified_Helmholtz}
    \left[ \nabla_T + n^2 (\textbf{r}) k_0^2 - \beta'^2 \left(1 + \frac{2x}{\rho}\xi(\textbf{r}) \right) \right] |\psi' \rangle = 0,
\end{equation}
where $|\psi'\rangle$ denotes the optical field in the curved fiber, and $\beta'$ is its associated propagation constant. The geometric factor $\xi(\textbf{r})$ accounts for the effect of material compression:
\begin{equation}
    \label{eqn:geometric_factor}
    \xi(\textbf{r}) = 1- \frac{n (\textbf{r}) - 1}{n (\textbf{r})} (1 - 2\sigma),
\end{equation}
where $\sigma$ is the Poisson ratio of the fiber material.

\subsection*{From modified Helmholtz equation to eigenvalue problem}

The modified Helmholtz equation can be reformulated as an eigenvalue problem. First, the field $|\psi'\rangle$ in the curved fiber can be expressed as a linear superposition of the eigenmodes of a straight fiber (i.e., in the limit $\rho \to \infty$):
\begin{equation}
    \label{eqn:mode_decompose}
    |\psi' \rangle = \sum_{i=1}^{N} {c_i |\psi_i \rangle},
\end{equation}
where $|\psi_i\rangle$ is the $i$-th eigenmode of the straight fiber with propagation constant $\beta_i$, satisfying:
\begin{equation}
    \label{eqn:helmholtz_straight}
    \left[ \nabla_T + n^2 (\textbf{r}) k_0^2 - \beta^2_i  \right] |\psi_i \rangle = 0,
\end{equation}
Substituting Equations~\ref{eqn:mode_decompose} and~\ref{eqn:helmholtz_straight} into the modified Helmholtz equation~\ref{eqn:modified_Helmholtz} gives:
\begin{equation}
    \label{eqn:eigen_derive_1}
    \sum_{i=1}^{M} {c_i \left[ \beta_i^2 - \beta'^2 \left(1 + \frac{2x}{\rho} \xi(\textbf{r}) \right) \right] |\psi_i \rangle } = 0,
\end{equation}
Projecting both sides onto the straight fiber eigenmode basis by left-multiplying with $\langle \psi_j|$, we obtain:
\begin{equation}
    \label{eqn:eigen_derive_2}
    c_j \beta_j^2 = \beta'^2 \left(c_j + \frac{2}{\rho} \sum_{i=1}^{N} {c_i \langle \psi_j | x\xi(\textbf{r}) | \psi_i \rangle } \right),
\end{equation}
Rearranging Equation~\ref{eqn:eigen_derive_2}, we obtain an eigenvalue problem expressed in the mode basis:
\begin{equation}
    \label{eqn:eigen_derive_3}
    \frac{1}{\beta_j^2} \left( c_j + \frac{2}{\rho} \sum_{i=1}^{N} {c_i \Gamma_{ji}} \right)= \frac{1}{\beta'^2} c_j, 
\end{equation}
where $\Gamma_{ji} = \langle \psi_j | x\xi(\textbf{r})| \psi_i \rangle$ represents the bending-induced coupling between fiber modes. Equation~\ref{eqn:eigen_derive_3} can be rewritten in matrix form:
\begin{equation}
    \label{eqn:matrix_form_1}
    \hat{B} \textbf{c} = \frac{1}{\beta'^2} \textbf{c}, 
\end{equation}
where $\textbf{c} = \left[ c_1, c_2, ... c_N \right]^T$ and $B_{ij} = \frac{1}{\beta_{i}^2} (\delta_{ij} + \frac{2}{\rho} \Gamma_{ij})$. 
In theory, the mode basis should include both guided and radiation (non-guided) modes to form a complete set. However, restricting the basis to guided modes is sufficient when the modes of interest are well below their cut-off frequencies.
Although the operator $\hat{B}$ is generally non-Hermitian (i.e., $\hat{B}^\dagger \neq \hat{B}$), it can be factorized into the product of two operators:
\begin{equation}
    \label{eqn:B_decompose}
    \hat{B} = \hat{M}^{-1} \hat{K},
\end{equation}
where
\begin{subequations}
    \label{eqn:M_and_K}
    \begin{align}
        M_{ij} &= \beta_i^2 \delta_{ij}, \\
        K_{ij} &= \delta_{ij} + \frac{2}{\rho} \Gamma_{ij},
    \end{align}
\end{subequations}
Note that $\hat{M}$ is diagonal and $\hat{K}$ is Hermitian, since $\Gamma_{ij}^* = \Gamma_{ji}$. Therefore, Equation~\ref{eqn:matrix_form_1} can be recast as a generalized eigenvalue problem with eigenvalues $\lambda = 1/\beta'^2$:
\begin{equation}
    \label{eqn:generalized_eigenproblem}
    \hat{K} \textbf{c} = \lambda \hat{M} \textbf{c},
\end{equation}
The Hermitian nature of $\hat{K}$ and the diagonal structure of $\hat{M}$ greatly simplify the application of perturbation theory, which we use in the next section to derive the dispersion relation.

\subsection*{Solving for dispersion coefficients via perturbation theory}

In the previous section, we reformulated the problem as a generalized eigenvalue problem, shown in Equation~\ref{eqn:generalized_eigenproblem}. To proceed, we first normalize the eigenvectors using the weighted inner product:
\begin{equation}
    \label{eqn:normalize}
    \textbf{c}^{\dagger} \hat{M} \textbf{c} = 1,
\end{equation}
This normalization is equivalent to weighting each mode by the square of its propagation constant, i.e., $\sum_{i=1}^{M} \beta_i^2 |c_i|^2 = 1$.

To investigate the dispersion relation, we consider small perturbations to $\hat{K}$, $\hat{M}$, $\lambda$, and $\textbf{c}$ due to frequency-dependent effects. We define the perturbative expansions as follows:
\begin{subequations}
    \label{eqn:perturbation_theory}
    \begin{align}
        \hat{K}_\epsilon &= \hat{K} + \epsilon \hat{K}^{(1)} + \epsilon^2 \hat{K}^{(2)} + \mathcal{O}(\epsilon^3), \\
        \hat{M}_\epsilon &= \hat{M} + \epsilon \hat{M}^{(1)} + \epsilon^2 \hat{M}^{(2)} + \mathcal{O}(\epsilon^3), \\
        \lambda_\epsilon &= \lambda + \epsilon \lambda^{(1)} + \epsilon^2 \lambda^{(2)} + \mathcal{O}(\epsilon^3), \\
        \textbf{c}_\epsilon &= \textbf{c} + \epsilon \textbf{c}^{(1)} + \epsilon^2 \textbf{c}^{(2)} + \mathcal{O}(\epsilon^3),
    \end{align}
\end{subequations}
Here, $\epsilon$ is a dummy variable that indexes the perturbation order. In the context of dispersion analysis, we set $\epsilon = \omega - \omega_0$ to obtain the Taylor expansion of all relevant quantities about the central frequency $\omega_0$.

\subsubsection*{First-order perturbation}

By inserting the expansions in Equation~\ref{eqn:perturbation_theory} into Equation~\ref{eqn:generalized_eigenproblem} and collecting all terms proportional to $\epsilon$, we obtain the first-order approximation:
\begin{equation}
    \label{eqn:1st_derive_1}
    \hat{K}^{(1)} \textbf{c} + \hat{K} \textbf{c}^{(1)} = \lambda^{(1)} \hat{M} \textbf{c} + \lambda \hat{M}^{(1)} \textbf{c} + \lambda \hat{M} \textbf{c}^{(1)},
\end{equation}
Next, we left-multiply both sides by $\textbf{c}^\dagger$ and use the identity $\textbf{c}^{\dagger} \hat{K} = \lambda \textbf{c}^{\dagger} \hat{M}$, which holds because both $\hat{K}$ and $\hat{M}$ are Hermitian:
\begin{equation}
    \label{eqn:1st_derive_2}
    \textbf{c}^{\dagger} \hat{K}^{(1)} \textbf{c} = \lambda^{(1)} \textbf{c}^{\dagger} \hat{M} \textbf{c} + \lambda \textbf{c}^{\dagger} \hat{M}^{(1)} \textbf{c},
\end{equation}
Using the normalization condition from Equation~\ref{eqn:normalize}, we simplify Equation~\ref{eqn:1st_derive_2} to obtain the first-order correction to the eigenvalue:
\begin{equation}
    \label{eqn:lambda1}
    \lambda^{(1)} = \textbf{c}^{\dagger} (\hat{K}^{(1)} - \lambda \hat{M}^{(1)}) \textbf{c},
\end{equation}
Considering the perturbation arising from a small frequency shift $\delta \omega = \omega - \omega_0 = \epsilon$, the first-order term $\lambda^{(1)}$ corresponds to the frequency derivative of the inverse squared propagation constant:
\begin{equation}
    \label{eqn:lambda1_to_beta1}
    \lambda^{(1)} = \frac{d \lambda}{d \omega} = \frac{d}{d\omega} \beta'^{-2} = -2 \beta'^{-3} \frac{d \beta'}{d \omega},
\end{equation}
The first-order dispersion at $\omega = \omega_0$ can therefore be expressed as:
\begin{equation}
    \label{eqn:beta1_analytic}
    \frac{d \beta'}{d \omega} = -\frac{\beta'^3}{2} \lambda^{(1)} = -\frac{\beta'^3}{2} \textbf{c}^{\dagger} (\hat{K}^{(1)} - \lambda \hat{M}^{(1)}) \textbf{c},
\end{equation}
where
\begin{subequations}
    \label{eqn:K1_and_M1}
    \begin{align}
        \hat{K}^{(1)} &= \frac{d}{d \omega} \hat{K}, \\
        \hat{M}^{(1)} &= \frac{d}{d \omega} \hat{M},
    \end{align}
\end{subequations}
The dispersion operators $\hat{K}^{(1)}$ and $\hat{M}^{(1)}$ can be computed from a single run of modal analysis in the straight fiber, making the overall computational cost low. The reciprocal of Equation~\ref{eqn:beta1_analytic} gives the group velocity of the eigenmode $\textbf{c}$ in the curved fiber.

On the other hand, the first-order correction to the eigenvector $\textbf{c}$ can be obtained by substituting Equation~\ref{eqn:lambda1} into Equation~\ref{eqn:1st_derive_2}:
\begin{equation}
    \label{eqn:c1_derive_1}
    (\hat{K} - \lambda \hat{M}) \textbf{c}^{(1)} = (\lambda^{(1)} \hat{M} + \lambda \hat{M}^{(1)} - \hat{K}^{(1)}) \textbf{c},
\end{equation}
From Equation~\ref{eqn:c1_derive_1}, we know that $\textbf{c}^{(1)}$ lies in the subspace orthogonal to the eigenvector $\textbf{c}$. We choose $\textbf{c}^{(1)}$ such that $\textbf{c}^{\dagger} \hat{M} \textbf{c}^{(1)} = 0$, i.e., $\textbf{c}^{(1)}$ is an orthogonal complement to $\textbf{c}$. Assume $(\mu_k, \textbf{x}_k)$ are the eigenvalue-eigenvector pairs of $\hat{K} - \lambda \hat{M}$.  Based on the completeness of the eigenvectors, we can decompose $\textbf{c}^{(1)}$ into a superposition of the eigenvectors $\textbf{x}_k$, i.e., $\textbf{c}^{(1)} = \sum_k {\alpha_k \textbf{x}_k}$, and arrive at the first-order correction to the eigenvector $\textbf{c}$:
\begin{equation}
    \label{eqn:c1_express_1}
    \textbf{c}^{(1)}_i = \sum_{k} { \frac{\textbf{x}_k^{\dagger} (\lambda^{(1)}_i \hat{M} + \lambda_i \hat{M}^{(1)} - \hat{K}^{(1)}) \textbf{c}_i}{\mu_k} \textbf{x}_k },
\end{equation}
If we follow the assumption that there are no evanescent or radiation modes involved, and the fiber eigenmodes form an orthogonal basis, then we can rewrite Equation~\ref{eqn:c1_express_1}:
\begin{equation}
    \label{eqn:c1_express_2}
    \textbf{c}^{(1)}_i = \sum_{k \neq i} { \frac{\textbf{c}_k^{\dagger} \hat{M} (\lambda^{(1)}_i \hat{M} + \lambda_i \hat{M}^{(1)} - \hat{K}^{(1)}) \textbf{c}_i}{\lambda_k} \textbf{c}_k },
\end{equation}
where the summation excludes \(k = i\) to maintain orthogonality.

\subsubsection*{Second-order perturbation}

Likewise, the second-order correction is obtained by substituting Equation~\ref{eqn:perturbation_theory} into Equation~\ref{eqn:generalized_eigenproblem} and collecting terms of order $\epsilon^2$:
\begin{subequations}
    \label{eqn:2nd_derive_1}
    \begin{align}
        LHS&: \hat{K} \textbf{c}^{(2)} + \hat{K}^{(1)} \textbf{c}^{(1)} + \hat{K}^{(2)} \textbf{c}, \\
        RHS&: \lambda \hat{M} \textbf{c}^{(2)} + \lambda \hat{M}^{(1)} \textbf{c}^{(1)} + \lambda \hat{M}^{(2)} \textbf{c} + \lambda^{(1)} \hat{M} \textbf{c}^{(1)} + \lambda^{(1)} \hat{M}^{(1)} \textbf{c} + \lambda^{(2)} \hat{M} \textbf{c},
    \end{align}
\end{subequations}
Following a similar approach as for the first-order perturbation, we left-multiply by $\textbf{c}^{\dagger}$ and rearrange to isolate the second-order eigenvalue correction, yielding:
\begin{equation}
    \label{eqn:lambda2_1}
    \lambda^{(2)} = (\textbf{c}^{\dagger} \hat{K} \textbf{c}^{(1)} + \textbf{c}^{\dagger} \hat{K}^{(2)} \textbf{c}) - (\lambda \textbf{c}^{\dagger} \hat{M}^{(1)} \textbf{c}^{(1)} + \lambda \textbf{c}^{\dagger} \hat{M}^{(2)} \textbf{c} + \lambda^{(1)} \textbf{c}^{\dagger} \hat{M} \textbf{c}^{(1)} + \lambda^{(1)} \textbf{c}^{\dagger} \hat{M}^{(1)} \textbf{c} ),
\end{equation}
Simplifying the above expression, we arrive at the final form for $\lambda^{(2)}$:
\begin{equation}
    \label{eqn:lambda2_final}
    \lambda^{(2)} = \textbf{c}^{\dagger} (\hat{K}^{(2)} - \lambda^{(1)} \hat{M}^{(1)} - \lambda \hat{M}^{(2)} ) \textbf{c} + \textbf{c}^{\dagger} ( \hat{K}^{(1)} - \lambda \hat{M}^{(1)} - \lambda^{(1)} \hat{M} ) \textbf{c}^{(1)},
\end{equation}
This second-order correction relates directly to the group velocity dispersion. To see this, consider the Taylor expansion of the propagation constant $\beta (\omega)$ about $\omega_0$:
\begin{equation}
    \label{eqn:taylor_expansion}
    \beta(\omega) = \beta(\omega_0) + (\omega-\omega_0) \frac{d \beta}{d \omega}(\omega_0) + \frac{1}{2} (\omega - \omega_0)^2 \frac{d^2 \beta}{d \omega^2} (\omega_0) + ...,
\end{equation}
By letting $\epsilon = \omega - \omega_0$, we can obtain:
\begin{equation}
    \label{eqn:lambda2_to_beta2}
    \lambda^{(2)} = \frac{1}{2} \frac{d^2 \lambda}{d \omega^2} = \frac{d}{d \omega} (-\beta'^{-3} \frac{d \beta'}{d \omega}) = \frac{1}{\beta'^4} \left[ 3 \left( \frac{d \beta'}{d \omega} \right)^2 - \beta' \frac{d^2 \beta'}{d \omega^2} \right]
\end{equation}
After rearranging Equation~\ref{eqn:lambda2_to_beta2}, we arrive at the expression for group velocity dispersion:
\begin{equation}
    \label{eqn:beta2_analytic}
    \frac{d^2 \beta'}{d \omega^2} = -\beta'^{3} \left[ \lambda^{(2)} -3\beta'^{-4} \left( \frac{d \beta'}{d \omega} \right)^2 \right] = -\beta'^{3} \left( \lambda^{(2)} -\frac{3}{4} \beta'^2 \left(\lambda^{(1)}\right)^2 \right),
\end{equation}
Equation~\ref{eqn:beta2_analytic} provides an analytical expression. The computational procedure consists of three steps:
\begin{enumerate}
    \item Perform modal analysis of a straight fiber up to second-order precision in the frequency domain to obtain $\hat{K}, \hat{K}^{(1)}, \hat{K}^{(2)}, \hat{M}, \hat{M}^{(1)}, \hat{M}^{(2)}$.
    \item Solve for the eigenvalue-eigenvector pairs $(\lambda_k, \textbf{c}_k)$ in a curved fiber with radius of curvature $\rho$ using the generalized eigenvalue problem (Equation~\ref{eqn:generalized_eigenproblem}).
    \item Use Equations~\ref{eqn:lambda1},~\ref{eqn:c1_express_2}, and~\ref{eqn:lambda2_final} to obtain the first- and second-order perturbation effects, which are essentially matrix multiplications.
\end{enumerate}

If we can assume the dispersion in the spatial profile in a weakly perturbed fiber is much weaker than that of the propagation constant, Equation~\ref{eqn:lambda2_final} can be further simplified with the following approximations:
\begin{enumerate}
    \item The dispersion in $\hat{K}$ is negligible, i.e., $\hat{K}^{(1)} = \hat{K}^{(2)} = 0$.
    \item The dispersion in the eigenvector $\textbf{c}$ is negligible, i.e., $\textbf{c}^{(1)} = 0$.
\end{enumerate}
The assumption of slowly varying spatial profiles is validated by explicitly computing the modal dispersion at different wavelengths and estimating its derivative via finite differences. We compare the spectral decompositions of the matrices $\lambda \hat{M}^{(1)}$ and $\hat{K}^{(1)}$ and find that the ratio between the smallest eigenvalue of $\lambda \hat{M}^{(1)}$ and the largest eigenvalue of $\hat{K}^{(1)}$, i.e., $\min {| \lambda_{\lambda \hat{M}^{(1)}} |} / \max {|\lambda_{\hat{K}^{(1)}}|}$, exceeds 6000 for $\rho = 1$~cm, thereby justifying the approximations. Hence, the first- and second-order corrections to the eigenvalue reduce to:
\begin{subequations}
    \label{eqn:simplified_lambda1_and_lambda2}
    \begin{align}
        \lambda^{(1)} &= -\lambda \textbf{c}^{\dagger} \hat{M}^{(1)} \textbf{c}, \\
        \lambda^{(2)} &= -\textbf{c}^{\dagger} (\lambda^{(1)} \hat{M}^{(1)} + \lambda \hat{M}^{(2)}) \textbf{c},
    \end{align}
\end{subequations}
Here, $\hat{M}^{(1)}$ and $\hat{M}^{(2)}$ can be directly computed by expanding the frequency dependence of the propagation constants of the straight fiber modes in Equation~\ref{eqn:M_and_K}:
\begin{subequations}
    \label{eqn:M1_and_M2}
    \begin{align}
        M_{ij}^{(1)} &= 2\beta_i \frac{d \beta_i}{d \omega} \delta_{ij}, \\
        M_{ij}^{(2)} &= \left[ \left( \frac{d \beta_i}{d \omega} \right)^2 + \beta_i \frac{d^2 \beta_i}{d \omega^2} \right] \delta_{ij},
    \end{align}
\end{subequations}
Combining Equation~\ref{eqn:simplified_lambda1_and_lambda2} and Equation~\ref{eqn:M1_and_M2}, we obtain:
\begin{subequations}
    \label{eqn:simplified_lambda1_and_lambda2_final}
    \begin{align}
        \lambda^{(1)} &= -2\lambda \sum_{i=1}^{N} {\beta_i \frac{d \beta_i}{d \omega} |c_i|^2}, \\
        \lambda^{(2)} &= - \sum_{i=1}^{N} { \left( \lambda^{(1)} M_{ii}^{(1)} + \lambda M_{ii}^{(2)} \right) |c_i|^2} \\
        &= - \sum_{i=1}^{N} {\left[ 2\lambda^{(1)}\beta_i \frac{d \beta_i}{d \omega} + \lambda \left( \left( \frac{d \beta_i}{d \omega} \right)^2 + \beta_i \frac{d^2 \beta_i}{d \omega^2} \right) \right] |c_i|^2 },
    \end{align}
\end{subequations}
As a sanity check, consider an eigenmode $(\lambda_k, \textbf{e}_k)$ in a straight fiber. Equation~\ref{eqn:simplified_lambda1_and_lambda2_final} should reduce to the known unperturbed dispersion relation. Given that $\lambda = \lambda_k = 1/\beta_k^2$ and $|c_k|^2 = 1/\beta_k^2$ by the normalization in Equation~\ref{eqn:normalize}, we have:
\begin{subequations}
    \label{eqn:sanity_check_straight}
    \begin{align}
        \lambda_k^{(1)} &= -2 \beta_k^{-3} \frac{d \beta_k}{d \omega}, \\
        \lambda_k^{(2)} &= \frac{1}{\beta_k^4} \left[ 3 \left( \frac{d \beta_k}{d \omega} \right)^2 - \beta_k \frac{d^2 \beta_k}{d \omega^2} \right],
    \end{align}
\end{subequations}
which is consistent with Equations~\ref{eqn:lambda1_to_beta1} and~\ref{eqn:lambda2_to_beta2}.

Therefore, Equation~\ref{eqn:simplified_lambda1_and_lambda2_final} shows that the dispersion relation of an eigenmode in a curved fiber is effectively a weighted superposition of the dispersion relations of the constituent modes in the straight fiber, where the weights are given by the modal power fractions $|c_i|^2$.

\section*{Supplementary Note 2: Negative axicon method}

\subsection*{Derivation of the negative axicon method for LP$_{0,m}$ mode matching}

Consider an axicon phase profile $t_{A}$ with apex angle $\alpha$ and refractive index $n$. It generates a Bessel-Gaussian beam at the focal plane of the axicon, with focal length approximately $f_a \approx 0.64 z_R$, where $z_R$ is the Rayleigh range of the Bessel-Gaussian beam. The field at the focal plane can be expressed as:
\begin{equation}
    \label{eqn:BG_beam_Fresnel}
    \psi_{B} (x,y) \propto e^{j\frac{k_0}{2f_a} (x^2 + y^2)} \iint \left[ G_i(\xi, \eta) t_{A}(\xi, \eta) e^{j\frac{k_0}{2f_a}(\xi^2 + \eta^2)} \right] e^{-j\frac{2\pi}{\lambda f_a}(x\xi+y\eta)} d\xi d\eta,
\end{equation}
where $G_i$ is the incident Gaussian beam and $t_{A}(\xi, \eta) = e^{-jk_a(\xi^2 + \eta^2)^{1/2} }$ is the axicon phase profile with phase ramp $k_a$. Equation~\ref{eqn:BG_beam_Fresnel} can be approximated by~\cite{Brzobohat2008}:
\begin{equation}
    \label{eqn:BG_beam}
    \psi_{B} (r) \approx J_0(\kappa r) e^{-\frac{r^2}{W_z^2}},
\end{equation}
where $\kappa = k_0 \alpha(n-1)$ is inversely proportional to the spacing between adjacent rings, and $W_z$ defines the $1/e$ amplitude radius of the beam. An ideal wavefront should generate an $M$-magnified Bessel beam at the fiber input that matches the target fiber mode.
\begin{equation}
    \label{eqn:ideal_wavefront}
    \psi_{in}(x,y) \propto \psi_{B} \left( \frac{x}{M}, \frac{y}{M} \right) \propto e^{j\frac{k_0}{2f_aM^2}(x^2 + y^2)} \iint \left[ G_i(\xi,\eta) t_{A}(\xi,\eta) e^{j\frac{k_0}{2f_a}(\xi^2 + \eta^2)} \right] e^{-j\frac{2\pi}{\lambda f_a M}(x\xi + y\eta)} d\xi d\eta,
\end{equation}
By backward propagating this beam to the back focal plane (BFP) of the in-coupling lens, the target optical field at the BFP is:
\begin{equation}
    \label{eqn:BG_beam_BFP_Fourier}
    \psi_{BFP} (\mu,\nu) = \mathcal{F}^{-1} \{\psi_{in}(x,y)\}_{f_x = \frac{\mu}{\lambda f_1}, f_y = \frac{\nu}{\lambda f_1}},
\end{equation}
where $\mathcal{F}^{-1}$ denotes the inverse Fourier transform.
After some derivation and rearrangement, we arrive at the following expression for the optical field at the BFP:
\begin{equation}
    \label{eqn:BG_beam_BFP_final}
    \psi_{BFP} (\mu,\nu) \propto \iint e^{-j\frac{k_0}{2f_a\gamma^2}\left[ (\mu-\xi)^2 + (\nu - \eta)^2 \right]} G_i \left(\frac{\xi}{\gamma}, \frac{\eta}{\gamma} \right) t_A \left( \frac{\xi}{\gamma}, \frac{\eta}{\gamma} \right) e^{j\frac{k_0}{2f_a\gamma^2}(\mu\xi + \nu\eta)} d\xi d\eta,
\end{equation}
where $\gamma = \frac{f_1}{f_a M}$. Equation~\ref{eqn:BG_beam_BFP_final} is exactly the Fresnel diffraction of the field $G_i\,t_A \, e^{j\frac{k_0}{2f_a\gamma^2}(\mu\xi + \nu\eta)}$ propagated over a distance $z = -f_a \gamma^2$. 
The first two terms represent the incident beam and the axicon phase, respectively, while the third term acts as a Fresnel lens phase with focal length $f_L = -f_a \gamma^2$. This result suggests that a perfect Bessel-Gaussian beam can be generated at the fiber input facet by satisfying the following conditions:
\begin{enumerate}
    \item Choose a negative focal length for the axicon phase, i.e. $f_a < 0$ such that the distance for Fresenl diffraction is positive. In other words, it's a negative axicon with a reverse linear radial phase ramp.
    \item Choose a positive focal length for the Fresnel lens that matches the distance between the SLM and the in-coupling lens, i.e. $f_L = -f_a \gamma^2 = -\frac{f_1^2}{f_a M} > 0$.
\end{enumerate}
Here, $M$ is a parameter used to optimize the final coupling efficiency given the incident beam waist $W_0$ and the fiber core radius $R_c$. The optimal value of $M$ can be determined numerically by maximizing the modal overlap between the Bessel-Gaussian beam and the target fiber mode. 
For example, for the LP$_{0,7}$ mode, the optimal normalized Gaussian width $M W_0 / R_c$ lies between 1.55 and 1.7, yielding a modal overlap of approximately 86\%, as shown in the Supplementary Figure~\ref{sfig:negative_axicon}.

\subsection*{Comparison with the binary phase plate method}

A similar analysis can be carried out for another popular method, the binary phase plate (BPP) method~\cite{Demas2015-HoM}. Consider a binary phase plate $b(\xi, \eta)$ that defines the ring structure of a quasi-Bessel beam at the SLM plane:
\begin{equation}
    \label{eqn:quasiB_SLM}
    \psi_{SLM} (\xi,\eta) = G_i(\xi,\eta) b(\xi,\eta) e^{-j\frac{k_0}{2f_L}(\xi^2 + \eta^2)},
\end{equation}
where $f_L$ is the focal length of an additional Fresnel lens phase that controls the beam size for mode matching. The optical field at the BFP of the in-coupling lens is:
\begin{equation}
    \label{eqn:quasiB_BFP}
    \psi_{BFP} (\mu, \nu) \propto e^{j\frac{k_0}{2z_b}(\mu^2 + \nu^2)} \iint \left[ G_i(\xi,\eta) b(\xi,\eta) e^{-j\frac{k_0}{2f_L}(\xi^2 + \eta^2)} e^{j\frac{k_0}{2z_b}(\xi^2 + \eta^2)} \right] e^{-j\frac{2\pi}{\lambda z_b}(\mu\xi + \nu\eta)} d\xi d\eta,
\end{equation}
where $z_b$ is the distance between the SLM and the BFP of the in-coupling lens. By setting $f_L = z_b$, the quadratic phase terms inside the integral cancel out:
\begin{equation}
    \label{eqn:quasiB_BFP_final}
    \psi_{BFP} (\mu,\nu) \propto e^{j\frac{k_0}{2z_b}(\mu^2 + \nu^2)} \iint G_i(\xi,\eta) b(\xi,\eta)  e^{-j\frac{2\pi}{\lambda z_b}(\mu\xi + \nu\eta)} d\xi d\eta,
\end{equation}
The magnification of the optical system that relays the quasi-Bessel beam to the fiber input facet is given by $M = f_1 / z_b = f_1 / f_L$.

Notably, Equation~\ref{eqn:quasiB_BFP_final} reveals a quadratic phase aberration $e^{j\frac{k_0}{2 z_b}(\mu^2 + \nu^2)}$ at the BFP of the in-coupling lens, which leads to a deviation from the ideal mode-matching condition. Numerical analysis in Supplementary Figure~\ref{sfig:negative_axicon} confirms a lower optimal modal overlap of 74\%, with the normalized Gaussian width $M W_0 / R_c$ confined to a narrower range between 0.78 and 0.82. In general, the negative axicon method exhibits superior performance and is less sensitive to the operating parameter $M$.

\newpage

\begin{figure}[h]
    \centering
    \includegraphics[width=0.95\linewidth]{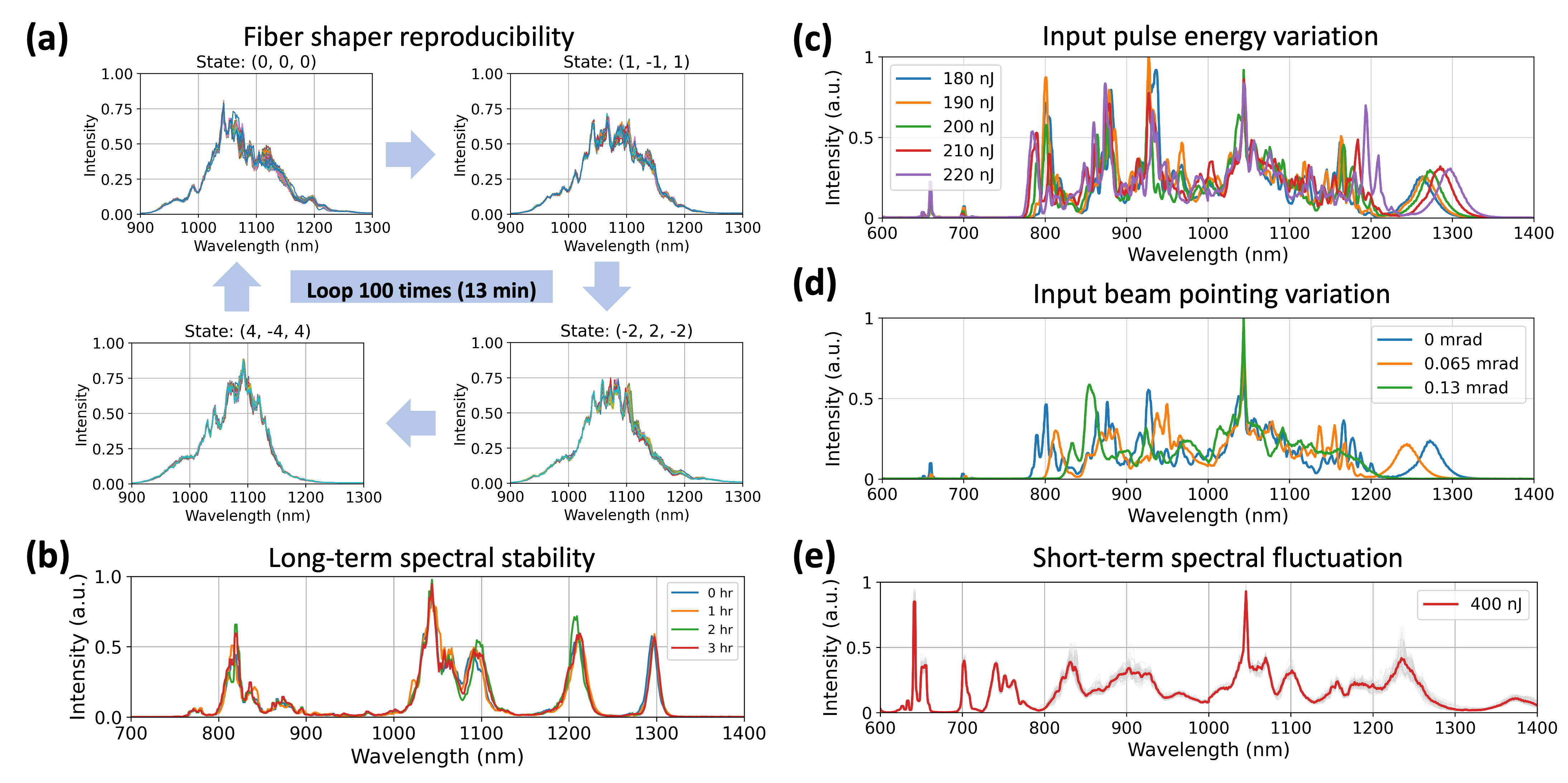}
    \caption{\textbf{System reproducibility and stability.} (a) Reproducibility of the second-generation fiber shapers. We performed a repeated-switching test across four distinct configurations, each defined by a unique set of three positions corresponding to the locations of the fiber shapers (in mm). For each configuration, 100 spectra were acquired over a 13-minute period—comparable to the duration of a single adaptive optimization run. The spectral consistency within each configuration was quantified using cosine similarity, yielding an average above 0.99 and a standard deviation below 0.0013.
    (b) Long-term spectral stability. A 9-meter 50-\textmu m-core SI-MMF was pumped by the LP$_{0,8}$ mode with an input pulse energy of 52~nJ. The positions of the fiber shapers were altered randomly, and the spectra were measured every one hour with the intial system configuration restored.
    (c) Spectral variation under fluctuating input pulse energy. The supercontinuum spectra in a 70-cm SI-MMF pumped by the LP$_{0,7}$ mode were measured under varying input pulse energies, with a deviation up to 10\% with respect to the baseline (200~nJ).
    (d) Spectral variation under fluctuating beam pointing. The supercontinuum spectra in a 70-cm SI-MMF pumped by the LP$_{0,7}$ mode were measured under varying input beam angles, with a deviation up to 0.013~mrad with respect to the normal incidence.
    (e) Short-term spectral stability. To quantify the fluctuation of the supercontinuum, we repeatedly measured the spectrum 100 times (gray-shaded traces) in a 70-cm SI-MMF pumped by the LP$_{0,7}$ mode with an input pulse energy of 400~nJ. Each measurement is an ensemble average over 40000 pulses, and the averaged cosine similarity is 0.985 with a standard deviation of 0.008.
    }
    \label{sfig:shaper_reproducibility}
\end{figure}

\begin{figure}[h!]
    \centering
    \includegraphics[width=0.85\linewidth]{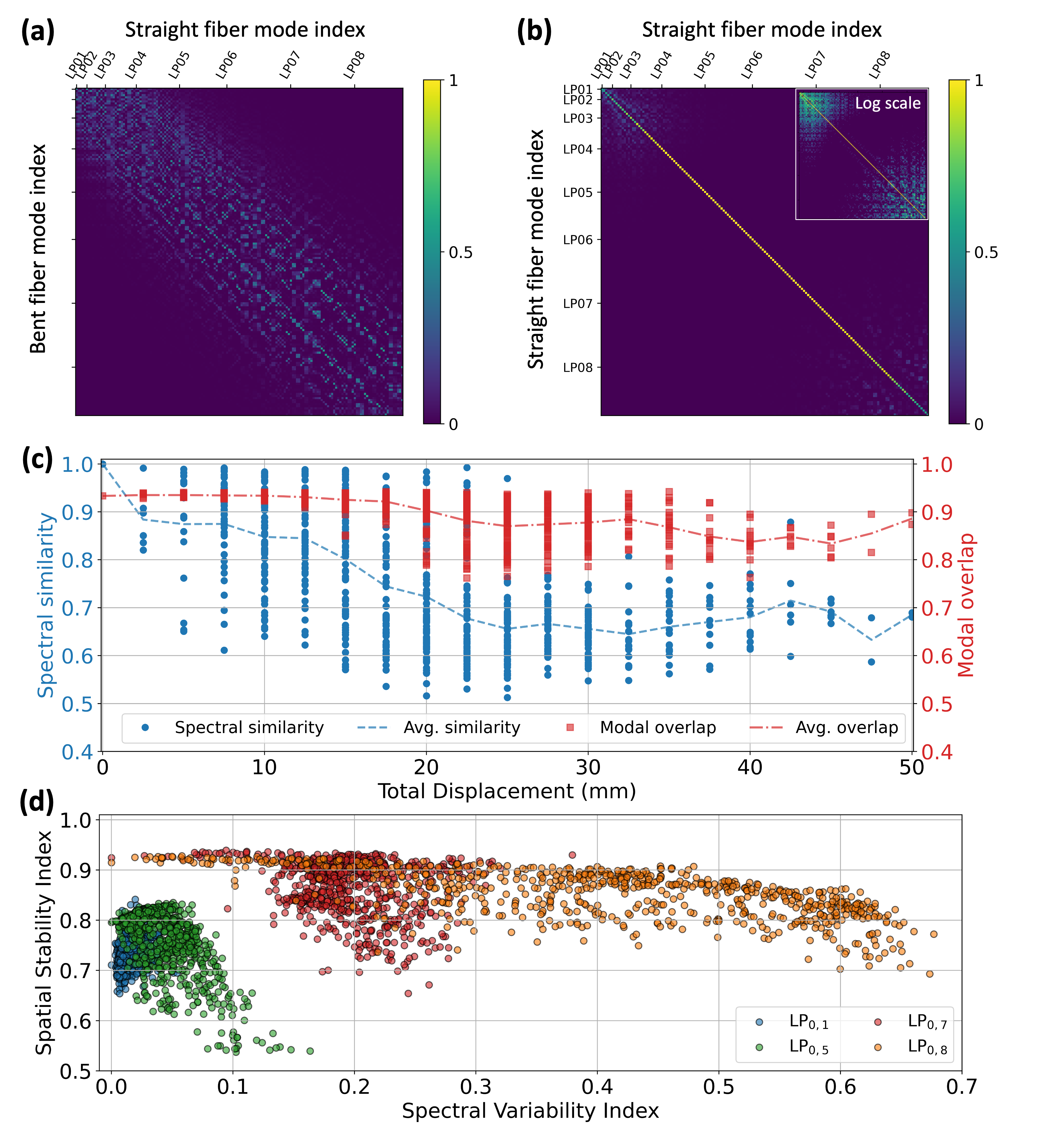}
    \caption{ \textbf{High-order modes are more resilient to bending.}
    (a) Scattering matrix $Q$ describing mode scattering from a straight fiber to a bent fiber. The matrix element $Q_{ij} = \langle \psi_i'(\mathbf{r}) \mid \psi_j(\mathbf{r}) \rangle$ is the coupling coefficient, where $\psi_j$ is the spatial mode profile of the $j$-th straight fiber mode, and $\psi_i'$ is that of the $i$-th bent fiber mode. The fiber modes are obtained by solving the modified Helmholtz equation in Equation~\ref{eqn:modified_Helmholtz}. The radius of curvature for the bent fiber is set to 1\,cm.
    (b) The matrix $T = Q^\dagger Q$ analyzes how the fiber mode composition changes after a short, locally bent fiber segment. The diagonal entries of $T$ represent the resilience of each fiber mode to mechanical bending. It is evident that high-order modes such as $\text{LP}_{0,7}$ exhibit stronger resilience compared to low-order modes.
    (c) The spectral similarity and the modal intensity overlap under 730 bending configurations displays spatial--spectral decorrelation. Bending is characterized by the total displacement. The spectral similarity is defined as the cosine similarity between the initial (unperturbed) spectrum and the spectra under different shaper configurations. The intensity modal overlap is defined as the cosine similarity between the measured beam intensity profile and the theoretical LP$_{0,7}$ mode.
    (d) Experimentally characterized spectral variability and spatial stability of different fiber modes. The spectral variability index is defined as 1 - $S$, where $S$ is the spectral similarity. The spatial stability is defined as the cosine similarity between the measured beam intensity profile and the target beam profile given by the launching condition. 730 fiber shaper configurations were examined for each launching mode.}
    \label{sfig:bent_fiber_modal_simulation}
\end{figure}

\begin{figure}[h]
    \centering
    \includegraphics[width=0.95\linewidth]{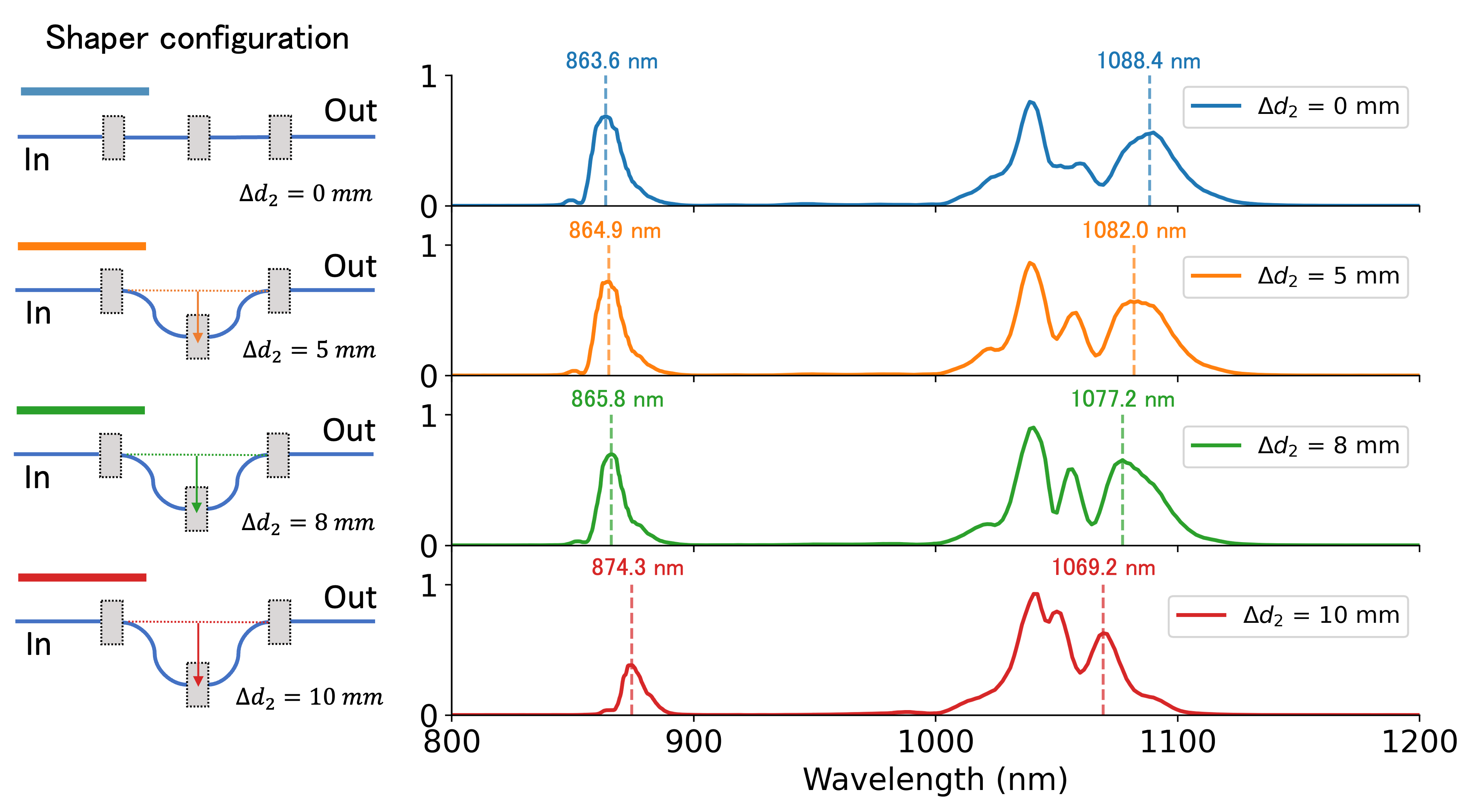}
    \caption{\textbf{Observation of bending-induced modified dispersion characteristics through wavelength shifts in dispersive wave generation.} The spectra are generated in a 70-cm 50-\textmu m-core SI-MMF with a LP$_{0,8}$ pump under different shaper configurations. A 16-nJ input pulse energy is applied, at which soliton formation and dispersive wave generation are predominant, while other nonlinear effects are not as pronounced due to their longer characteristic lengths. The spectra show wavelength shifts of solitons and dispersive waves as a result of modified dispersion properties that render the phase-matching condition.
    }
    \label{sfig:dwg}
\end{figure}

\begin{figure}[h]
    \centering
    \includegraphics[width=0.7\linewidth]{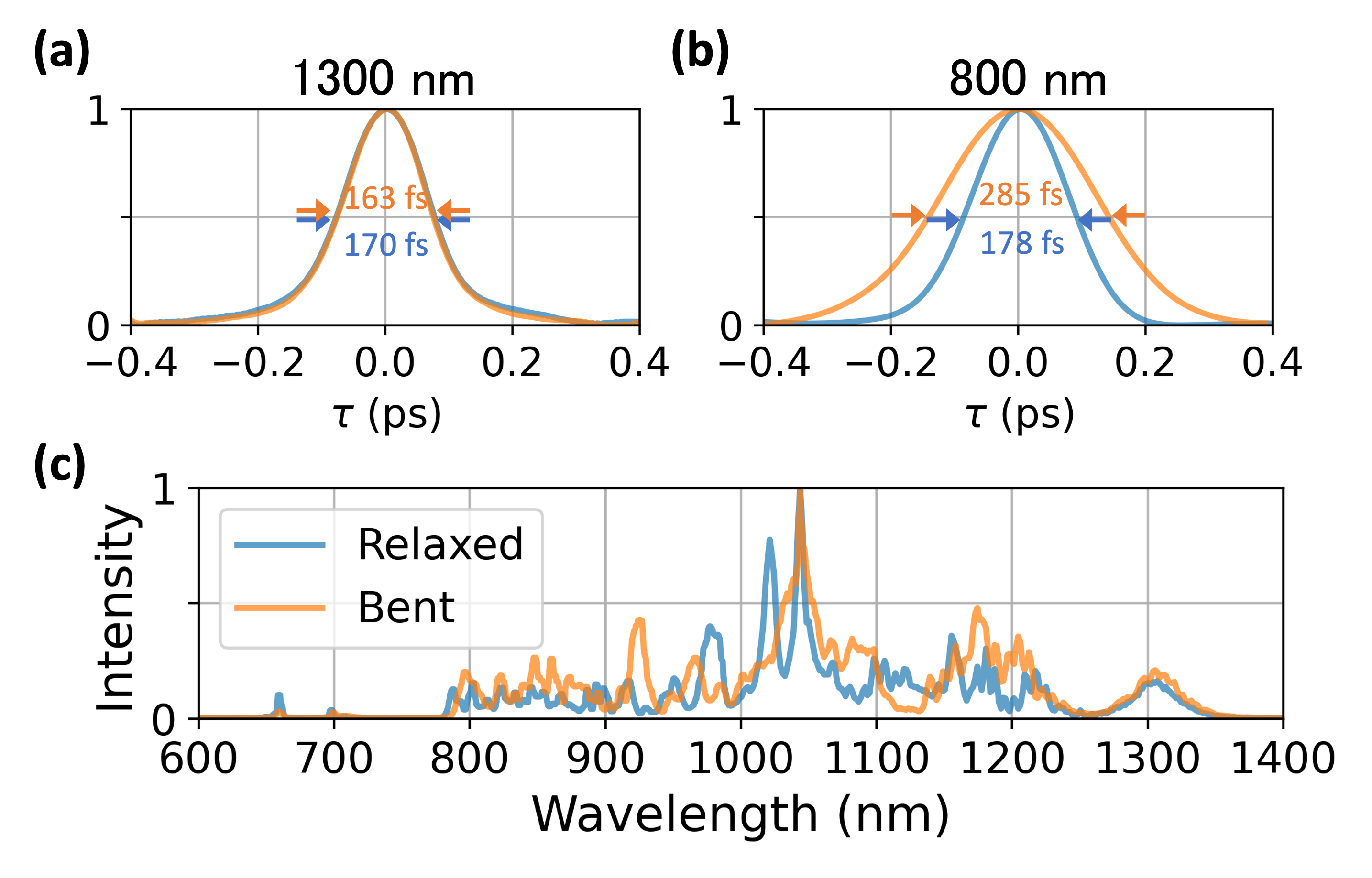}
    \caption{\textbf{Pulse duration characterization under bending.}
    Autocorrelation measurement of solitons at 1300~nm (a) and dispersive waves at 800~nm (b). (c) The corresponding spectra in a relaxed and a bent SI-MMF.}
    \label{sfig:pulse_duration}
\end{figure}

\begin{figure}[h]
    \centering
    \includegraphics[width=0.95\linewidth]{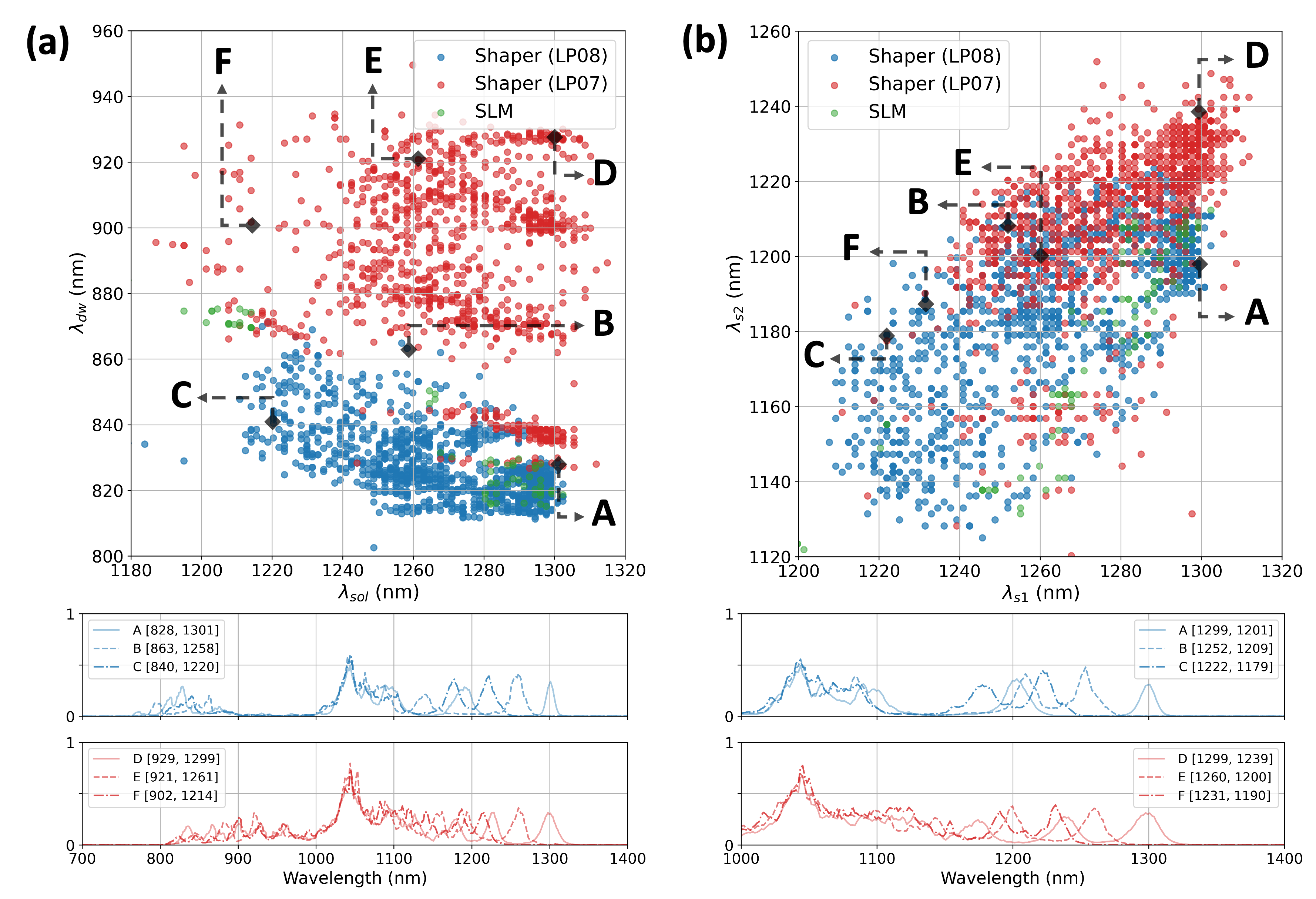}
    \caption{ \textbf{Decorrelation between supercontinuum components.}
    (a) Soliton–dispersive wave decorrelation and (b) multi-soliton decorrelation. The scatter plots show the achievable wavelength combinations for soliton–dispersive wave (a) and soliton–soliton (b) tuning. Selected example spectra are shown below. Due to spectral decorrelation, mechanical perturbations via fiber shapers enable flexible dual-wavelength tunability in both cases.}
    \label{sfig:spectral_decorrelation}
\end{figure}

\begin{figure}[h]
    \centering
    \includegraphics[width=0.95\linewidth]{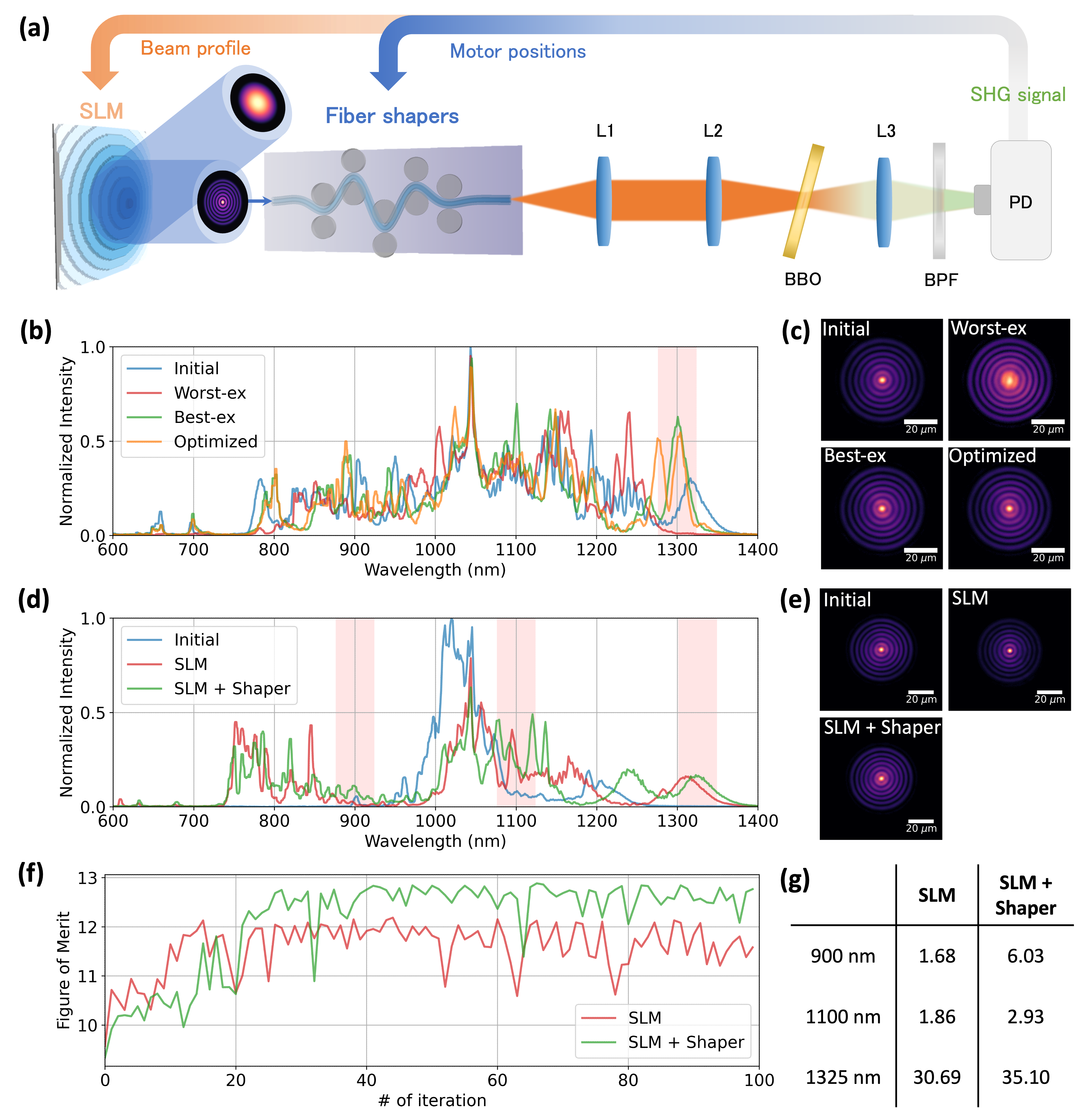}
    \caption{\textbf{Adaptive pulse optimization.}
    (a) System schematic of adaptive pulse optimization using SHG feedback. 
    (b–c) Single-wavelength optimization using the SHG signal as the figure of merit for 1300~nm. (b) Four spectra corresponding to different fiber configurations: (1) the unperturbed fiber (Initial), (2) the worst case in the exhaustive search (Worst-ex), (3) the best case in the exhaustive search (Best-ex), and (4) the adaptive optimization result (Optimized). (c) The corresponding spatial profiles. 
    (d–g) Multi-band optimization using the logarithmic sum of spectral density as the figure of merit. (d) The spectra before (Initial) and after optimization (SLM and SLM~+~Shaper). The initial state corresponds to an LP$_{0,6}$ launching mode and an unperturbed fiber. For the SLM-only optimization, two phase mask parameters—the axicon apex angle and the Fresnel lens focal length—are optimized. In the SLM~+~Shaper case, motor positions are jointly optimized with the SLM parameters. (e) The corresponding spatial profiles. (f) Optimization curves. (g) Enhancement ratio for each spectral band compared to the initial state.
    }
    \label{sfig:adaptive_optimization}
\end{figure}

\begin{figure}[h]
    \centering
    \includegraphics[width=1.00\linewidth]{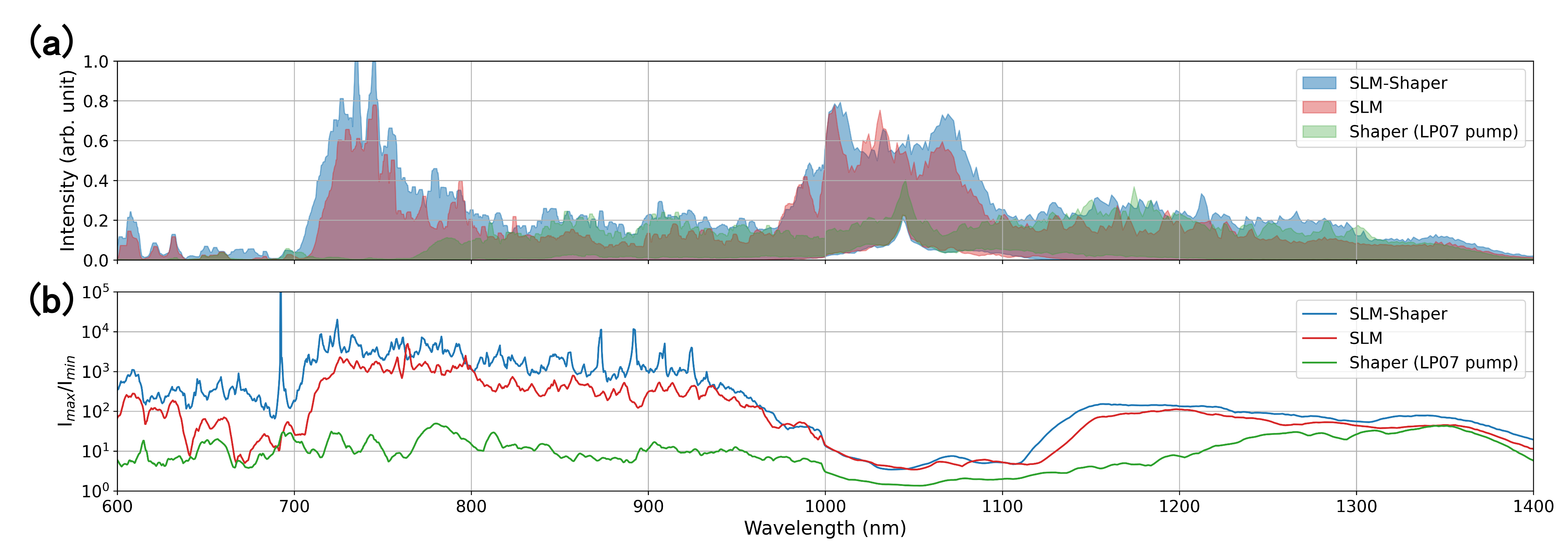}
    \caption{\textbf{Spectral tunability enhancement via integrating wavefront shaping and mechanical bending.}
    (a) Spectral dynamic range attained by three different approaches: (1) SLM~+~Shaper (blue), (2) SLM-only (red), and (3) Shaper-only (with a fixed LP$_{0,7}$ pump). An exhaustive search was performed for each case with 6351, 800, and 730 spectra, respectively. The parameter space for the SLM includes any rotationally symmetric mode between LP$_{0,1}$ and LP$_{0,9}$, and the parameter space for the fiber shapers includes the positions of three shaper motors with a $\pm$1~cm travel range and a step size of 0.25~cm.
    (b) The corresponding spectral tunability of the three approaches, defined by the ratio between the maximum and minimum intensity ($I_\mathrm{max} / I_\mathrm{min}$).
    }
    \label{sfig:spectral_tunability}
\end{figure}

\begin{figure}[h]
    \centering
    \includegraphics[width=0.95\linewidth]{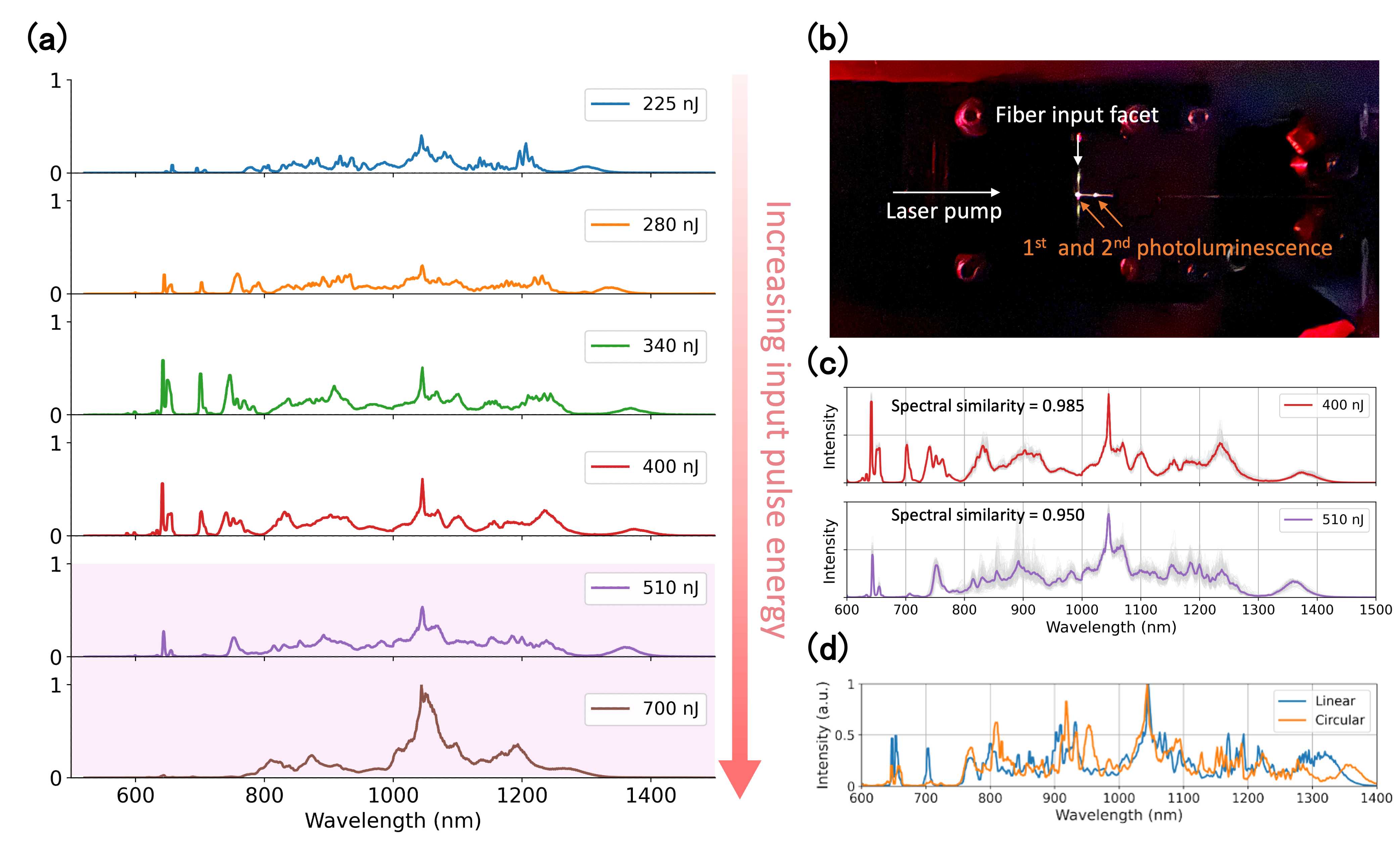}
    \caption{
    \textbf{Damage threshold analysis for LP$_{0,7}$ pump.} (a) When the input pulse energy increases to 510~nJ, the spectral broadening ceases, indicating the onset of laser-induced damage. Further increasing the pulse energy to 700~nJ results in a noticeable spectral collapse within 30 seconds, suggesting permanent damage that cannot be reversed by reducing the irradiance. This phenomenon is accompanied by self-focusing-incuded damage, as evidenced by the photoluminescence shown in (b). The first photoluminescent spot is close to the input facet, and the second is approximately 0.5~mm away. Moreover, a considerable increase in spectral fluctuation occurs when the input pulse energy is beyond 400~nJ, as shown in (c). Consequently, the input pulse energy is constrained below 400~nJ to ensure safe operation. (d) Supercontinuum spectra of linearly polarized and circularly polarized LP$_{0,7}$ mode pumps. The pump pulse energies are 280 and 400~nJ, respectively.
    }
    \label{sfig:damage_threshold}
\end{figure}

\begin{figure}[h]
    \centering
    \includegraphics[width=0.85\linewidth]{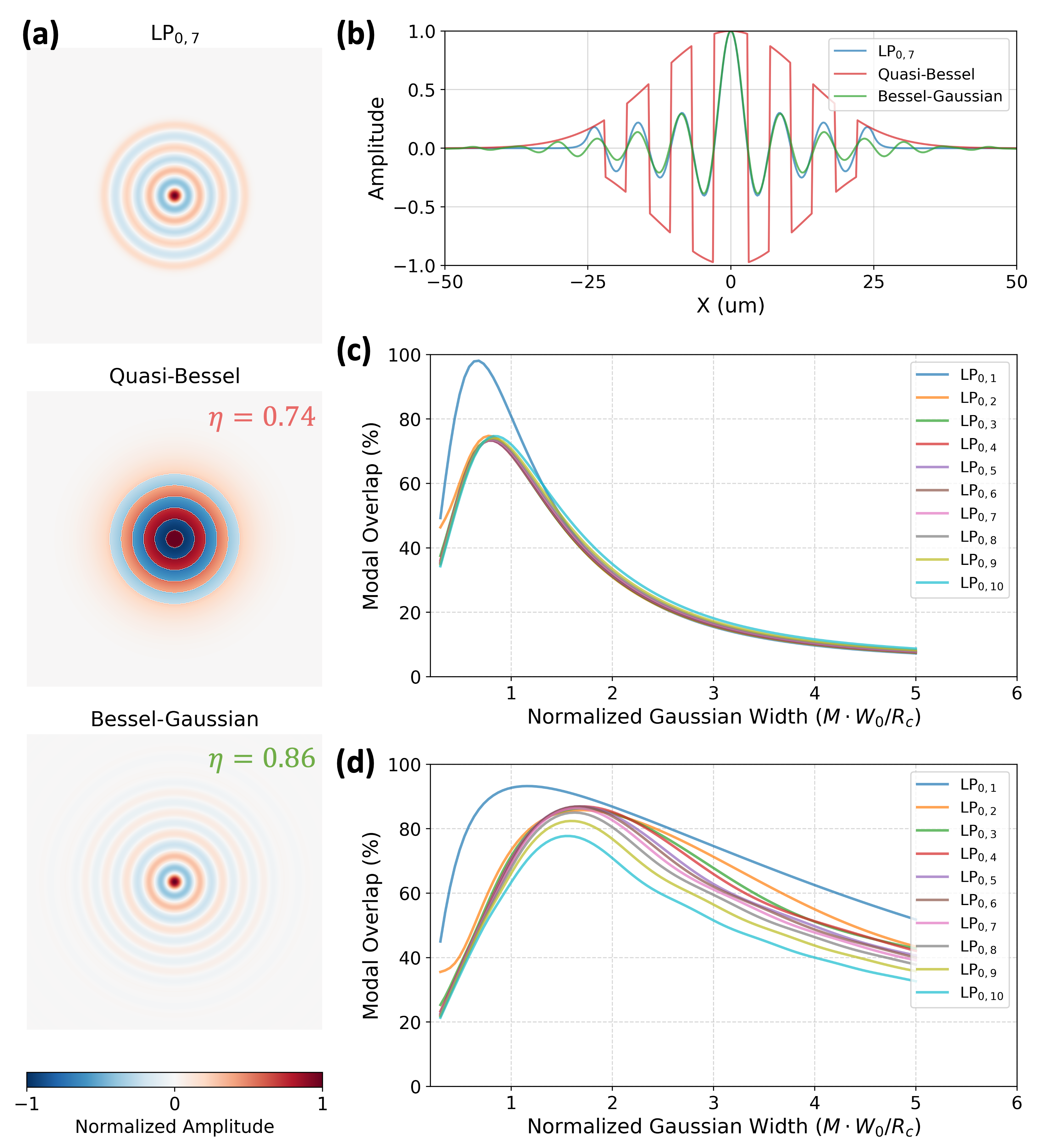}
    \caption{\textbf{Comparison between the BPP method and the negative axicon method}
    (a) Spatial profiles of the Bessel beam generated by two different approaches: (1) quasi-Bessel beam via the BPP method and (2) Bessel-Gaussian beam via the negative axicon method. The optimal coupling efficiencies $\eta$ are 0.74 and 0.86, respectively, for the LP$_{0,7}$ mode. The corresponding cross-sectional profiles are shown in (b). (c--d) Modal overlap analysis as a function of normalized Gaussian width for (c) the BPP method and (d) the negative axicon method. The numerical treatment follows Supplementary Note 2.
    }
    \label{sfig:negative_axicon}
\end{figure}

